\documentclass[showpacs,twocolumn,superscriptaddress]{revtex4}
\usepackage{doi}
\usepackage{hyperref}
\hypersetup{
  colorlinks=true,        
  linkcolor=blue,         
  citecolor=cyan,         
}

\usepackage{graphicx}
\usepackage{dcolumn}
\usepackage{bm}
\usepackage{color}
\usepackage{enumitem}
\usepackage{amsmath}
\usepackage{amssymb}
\usepackage{orcidlink}
\usepackage{subcaption}
\usepackage{graphicx} 
\begin{document}
\title{Gravitational Wave Signatures of Schwarzschild Black Hole in a Generalized Dehnen-Type $(1,4,\gamma)$ Dark Matter Halo}

\author{Bahromjon Shokirov}
\email{bahromshokirov05@gmail.com}
\affiliation{National University of Uzbekistan, Tashkent 100174, Uzbekistan }

\author{Arabboy Mirzakulov}
\email{arabboymiraqulov777@gmail.com}
\affiliation{National University of Uzbekistan, Tashkent 100174, Uzbekistan }

\author{Tursunali Xamidov}
\email{xamidovtursunali@gmail.com}
\affiliation{Institute for Theoretical Physics and Cosmology,
Zhejiang University of Technology, Hangzhou 310023, China}
\affiliation{Institute of Fundamental and Applied Research, National Research University TIIAME, Kori Niyoziy 39, Tashkent 100000, Uzbekistan}

\author{Sanjar Shaymatov}
\email{sanjar@astrin.uz}
\affiliation{Institute of Fundamental and Applied Research, National Research University TIIAME, Kori Niyoziy 39, Tashkent 100000, Uzbekistan}
\affiliation{University of Tashkent for Applied Sciences, Str. Gavhar 1, Tashkent 100149, Uzbekistan}

\date{\today}

\begin{abstract}
In this paper, we investigate timelike geodesic motion, periodic orbits, and the associated gravitational-wave signals around a Schwarzschild-like black hole (BH) embedded in a generalized Dehnen-type dark matter (DM) halo. We show that the Dehnen-type $(1,4,\gamma)$ DM halo profile modifies test-particle dynamics, with increasing the parameter of density profile, $\gamma$, leading to larger marginally bound orbit (MBO) and innermost stable circular orbit (ISCO) radii and angular momenta, together with a higher ISCO energy. These findings provide further insight into the role of the DM distribution in modifying the orbital dynamics, energy, and angular momentum of timelike test particles near the BH. Furthermore, we investigate the gravitational-wave signals produced by a stellar-mass compact object moving along periodic orbits around a supermassive BH embedded in a generalized Dehnen-type DM halo. Using the numerical kludge approach, we calculate the orbital trajectories and the corresponding gravitational-wave polarizations. We find that increasing the halo parameters $\gamma$, $\rho_s$, and $r_s$ produces larger periodic orbits, longer orbital periods, and lower waveform amplitudes. The resulting spectra lie mainly in the millihertz frequency range, while several characteristic-strain peaks lie above the sensitivity curves of future space-based gravitational-wave detectors such as LISA, Taiji, and TianQin. These results suggest that the surrounding DM halo may leave observable imprints on extreme mass-ratio inspiral (EMRI) gravitational-wave signals.

\end{abstract}

\maketitle

\section{Introduction}

Black holes (BHs), predicted by General Relativity (GR) as exact solutions of Einstein's equations \cite{1916SPAW.......189S,2015arXiv151202061B,1963PhRvL..11..237K}, have been observationally confirmed through LIGO-VIRGO gravitational-wave (GW) observations \cite{Abbott16a,Abbott16b} and the imaging of supermassive BHs in M87* and Sgr A*~\cite{Akiyama19L1,Akiyama19L6,Akiyama22L12,EHT2022L14}. 
These observations establish BHs as unique laboratories for testing gravity and probing fundamental physics. Nevertheless, important questions remain regarding the interaction of BHs with their environments, particularly the nature and distribution of dark matter (DM). Understanding the role of DM in BH spacetimes is therefore an essential task and remains the main problem in GR. Consequently, further studies and precision tests of the DM halo effects are crucial to investigate GW phenomena, the structure of the BH horizon, and gravity in the strong-field regime \cite{Will14LRR, Psaltis20PRL}.

In many astrophysical settings, supermassive BHs at the centers of galaxies are generally to be embedded in matter distributions containing DM halos. Current cosmological observations indicate that baryonic matter constitutes only about 5\% of the Universe, whereas DM and dark energy contribute approximately 27\% and 68\%, respectively. Supermassive BHs can be found in complex and highly dynamical environments and are widely believed to power active galactic nuclei, with growing evidence suggesting that they are surrounded by DM halos~\cite{Iocco15NatPhy,Bertone18Nature}. Furthermore, DM plays a crucial role in explaining galactic rotation curves, the dynamics of galaxy clusters, and the formation of large-scale cosmic structures~\cite{Rubin70ApJ,Corbelli00MNRAS,Davis85ApJ}. Although the fundamental nature of DM remains unknown and its interactions appear to be primarily gravitational, a wide range of observational evidence supports its existence \cite{Bertone05,deSwart17Nat,Wechsler18}.

It is also worth noting that BHs embedded in DM halos have emerged as promising laboratories for investigating the nature of DM and its signatures in GW physics. This has led to the development of numerous BH solutions that incorporate different DM halo density profiles~\cite{Merritt06ApJ,Dutton_2014,Burkert95ApJL,Dehnen93,Shukirgaliyev21A&A,Gohain24DM,Pantig22JCAP,Al-Badawi25JCAP,Uktamov25EPJC,Wang2025PhRvD}. In addition, several alternative models have also been considered, including DM halos sourced by phantom scalar fields \cite{Li-Yang12,Shaymatov21d,Shaymatov21pdu,Shaymatov22a,Hou18-dm} and analytical supermassive BH solutions within DM environments \cite{Cardoso22DM,Shen24PLB,Shen25PLB,BoWang2025JCAP...01..086L}. As mentioned, a DM halo surrounding a BH influences not only the dynamics of test particles but also the underlying spacetime geometry, thereby modifying the horizon structure and affect observable quantities, e.g., the innermost stable circular orbits (ISCOs), BH shadows, and GW signals. Some of these effects have been extensively studied within the Dehnen-type halo framework. In particular, BH solutions with Dehnen-type DM halos have been considered to model ultra-faint dwarf galaxies~\cite{Pantig22JCAP}. More recently, BH spacetimes with various Dehnen-type density profiles have been investigated, revealing their impact on BH observables~\cite{Gohain24DM,Al-Badawi25JCAP,Uktamov25EPJC,Al-Badawi25CPC,Al-Badawi25CTP_DM,Alloqulov25EPJC_GW1,Xamidov25PDU,BoWang2026Univ...12...48C,Xamidov25EPJC...85.1193X}.

GW observations and analysis also provide a promising framework for investigating the effects of DM halos on BH dynamics and for testing gravity in the strong-field regime. Future space-based detectors, including LISA~\cite{Amaro-Seoane2017LISA} and Taiji~\cite{10.1093/nsr/nwx116}, will probe a broad range of sources, particularly stellar-mass binaries and extreme mass-ratio inspirals (EMRIs) that consist of a compact object of stellar-mass orbiting a supermassive BH ~\cite{Hughes_2001,Amaro-Seoane18LRR,Babak17PRD}, allowing precision studies of BH–DM interactions. EMRIs generate long-lived low-frequency GWs that characterize detailed information about the orbital dynamics and the underlying spacetime geometry. Modifications to background spacetime through BH–DM interactions alter orbital frequencies, resonance conditions, and stability of circular orbits, thus affecting zoom–whirl dynamics and GW signals emitted ~\cite{Glampedakis02PRD,Ruangsri14PRD}. Consequently, EMRIs provide an exceptionally sensitive probe of BH horizon structure, gravity in the strong-field regime~\cite{Gair13CQG,Barack19CQG} and the surrounding astrophysical environment, including DM halos around supermassive BHs~\cite{Yue19ApJ,Duque24PRL,Dai24PRD}.

Periodic orbits provide a natural framework for describing the dynamics of EMRIs that generate gravitational wave signals~\cite{Levin_2008,Grossman_2009,Misra_2010}. They arise when the radial ($\omega_r$) and azimuthal ($\omega_\varphi$) motions satisfy a resonance condition (\(r\)-\(\varphi\)), causing the orbiting compact object to return to its initial position after a finite number of oscillations. These resonant trajectories exhibit the characteristic zoom--whirl behavior and are uniquely specified by the rational frequency ratio $\omega_\varphi/\omega_r$ and the integers $(z,w,v)$, corresponding to the zoom, whirl, and vertex numbers~\cite{Levin_2008,Levin_2009}.
This periodic-orbit formalism provides a unified framework for investigating GW emission, zoom--whirl dynamics, and resonance structures in EMRIs~\cite{Glampedakis02PRD}. Originally developed for Schwarzschild and Kerr spacetimes using topological integers $(z,w,v)$~\cite{Levin_2008,Levin_2009,Bambhaniya20,Rana19}, it has since been widely applied to a variety of BH geometries to explore periodic orbits \cite{Healy09PRL,Levin2010,Pugliese13,Babar17PRD,Liu18,Lin23,Yao23,Chan25,Lin22,Tu23,Deng20,Wei19,Zhang22,JIANG2024,Wei25,Alloqulov26GW1,Sharipov25} and their gravitational-wave signatures~\cite{Barausse14PRD,Cardoso22PRD,Ahmed25GW2,Yang24JCAP,Shabbir25,Junior24,Yang24,Haroon25,Lu25GW,Chen25,Li25,Ahmed26GW1,Wang25JCAP,Zhang26EPJC}. Therefore, EMRI gravitational-wave signals may provide a powerful probe of the DM distribution surrounding supermassive BHs~\cite{Navarro96ApJ,Gondolo99PRL}.

In this paper, we extend the study of periodic orbits around a Schwarzschild-like BH embedded in a generalized Dehnen-type DM halo and investigate whether the DM halo leaves observable imprints on GWs emitted by EMRI systems. We first derive the equations of timelike geodesic motion using the Lagrangian formalism and analyze the influence of the DM halo on the effective potential. We then examine the MBO and the ISCO, together with the dependence of the rational number $q$ on the test-particle energy and orbital angular momentum for different DM halo parameters. Periodic trajectories are classified by the integers $(z,w,v)$ and used to study, within the numerical kludge framework, how the surrounding DM halo modifies the gravitational-wave signals produced by a stellar-mass compact object orbiting the central supermassive BH.

The paper is organized as follows. Section~\ref{sec2} introduces the Schwarzschild-like spacetime surrounded by a generalized Dehnen-type DM halo and examines the timelike geodesics, effective potential, MBO, and ISCO. In Section~\ref{sec3}, we study periodic orbits and determine how the halo parameters affect the rational number $q$ and the corresponding energy and angular momentum. Section~\ref{sec4} focuses on the gravitational-wave signals produced by periodic orbits in the EMRI system, calculated using the numerical kludge method. The waveform, frequency spectrum, and characteristic strain are also discussed. The main conclusions are presented in Section~\ref{summary}.

\section{Spacetime metric and timelike geodesics}\label{sec2}

In this section, we analyze the effects of DM on timelike geodesics by considering a recently derived fully generalized Schwarzschild-like BH solution embedded in a Dehnen-type DM halo \cite{boltaev2026arxiv}. The corresponding spacetime metric is given by
\begin{equation}
ds^{2}=-f(r)dt^{2}+f(r)^{-1}dr^{2}
+r^{2}\left(d\theta^{2}+\sin^{2}\theta d\phi^{2}\right),
\end{equation}
where
\begin{equation}
    f(r)=1-\frac{2M}{r}-\frac{8\pi\rho_s r_s^3}{(3-\gamma)r}
    \left(\frac{r}{r+r_s}\right)^{3-\gamma}.
\end{equation}
Here, $M$ denotes the BH mass, while $\rho_s$ and $r_s$ are the characteristic density and scale radius of the Dehnen-type DM halo, respectively. The parameter $\gamma$ controls the shape of the density profile and lies in the range $0 \leq \gamma < 3$.

\begin{figure}[!htb]
    \centering
    \includegraphics[scale=0.38]{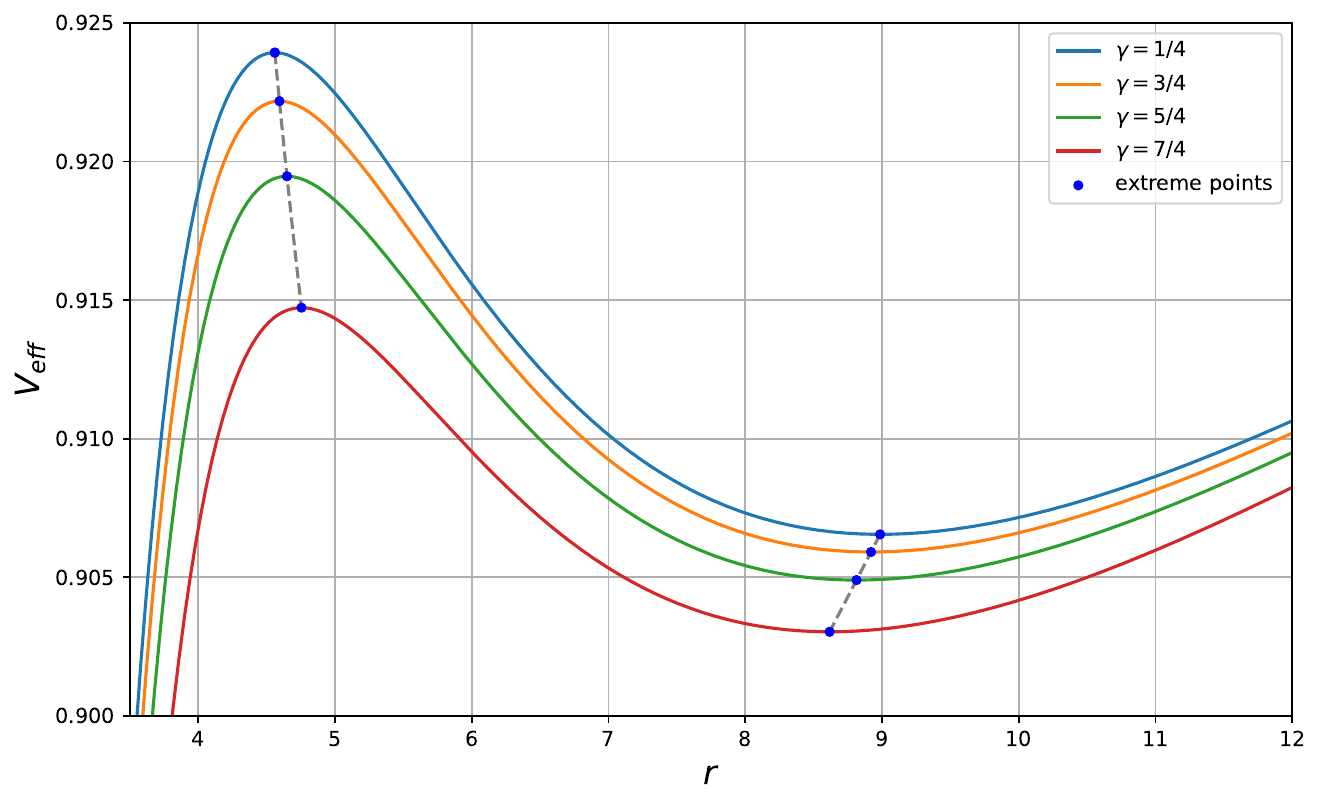}
    \caption{Radial dependence of the effective potential for the generalized Schwarzschild-like BH embedded in a Dehnen-type DM halo, shown for different values of the halo parameter $\gamma$. The remaining parameters are fixed at $\rho_s=0.3$, $r_s=0.2$, and $L=3.7$. The marked points indicate the extrema of the effective potential.
 }
    \label{fig:effpotential}
\end{figure}

To describe the motion of a test particle in this spacetime, we use the Lagrangian formalism. The corresponding Lagrangian is given by ~\cite{1983mtbh.book.....C}
\begin{equation}
\mathcal{L}=\frac{1}{2}m\,g_{\mu\nu}\,\frac{dx^{\mu}}{d\tau}\,\frac{dx^{\nu}}{d\tau},
\end{equation}
where $\tau$ denotes the proper time and $m$ is the mass of the test particle. For simplicity, we set $m=1$. The corresponding generalized momentum is then given by
\begin{equation}
    p_{\mu}=\frac{\partial {\cal L}}{\partial \dot{x}^{\mu}}=g_{\mu \nu}\dot{x}^{\nu}\, .
\end{equation}
Since the spacetime is independent of the coordinates $t$ and $\phi$, the corresponding energy $E$ and angular momentum $L$ are conserved and can be written as
\begin{equation}
     -\left(1-\frac{2M}{r}-\frac{8\pi\rho_sr_s^3}{(3-\gamma)r}\Big(\frac{r}{r+r_s}\Big)^{3-\gamma}\right)\dot{t}=-E
\end{equation}
\begin{equation} \label{eq:azimthal}
    r^{2}\sin^{2}\theta\dot{\phi}=L
\end{equation}
Using the conservation laws and the normalization condition
$g_{\mu\nu}u^{\mu}u^{\nu}=-1$, where $u^\mu=dx^\mu/d\tau$ is the four-velocity, the radial equation of motion for a test particle moving in the equatorial plane can be written as
\begin{equation}\label{eq:radialeq}
\dot{r}^{2}=E^{2}-V_{\rm eff}(r),
\end{equation}
where the effective potential $V_{\rm eff}(r)$ is given by
\begin{equation}
V_{\rm eff}(r)=\left[1-\frac{2M}{r}-\frac{8\pi\rho_s r_s^3}{(3-\gamma)r}
    \left(\frac{r}{r+r_s}\right)^{3-\gamma}\right]\left(1+\frac{L^{2}}{r^{2}}\right).
\end{equation}
The effective potential provides an important tool for analyzing the motion of test particles. Therefore, the effect of the Dehnen-type $(1,4,\gamma)$ DM halo parameter $\gamma$ on particle dynamics is examined through the effective potential, as shown in Fig.~\ref{fig:effpotential}. Increasing $\gamma$ shifts the unstable circular orbit, corresponding to the maximum of the effective potential, toward smaller radial distances. In contrast, the stable circular orbit, associated with the minimum of the effective potential, moves outward. This behavior indicates that the DM parameter $\gamma$ modifies the effective potential and, consequently, the orbital configurations of test particles.
\begin{figure*}[!htb]
    \centering
   
    \includegraphics[width=0.32\textwidth]{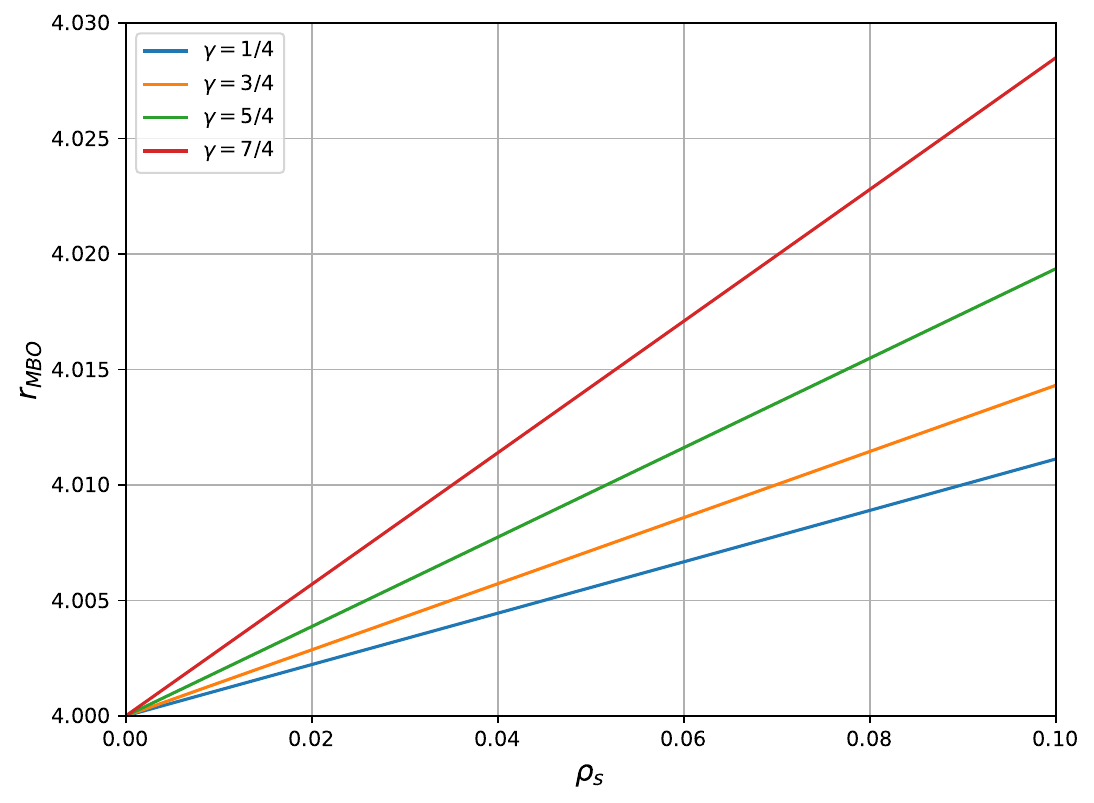}
    \includegraphics[width=0.32\textwidth]{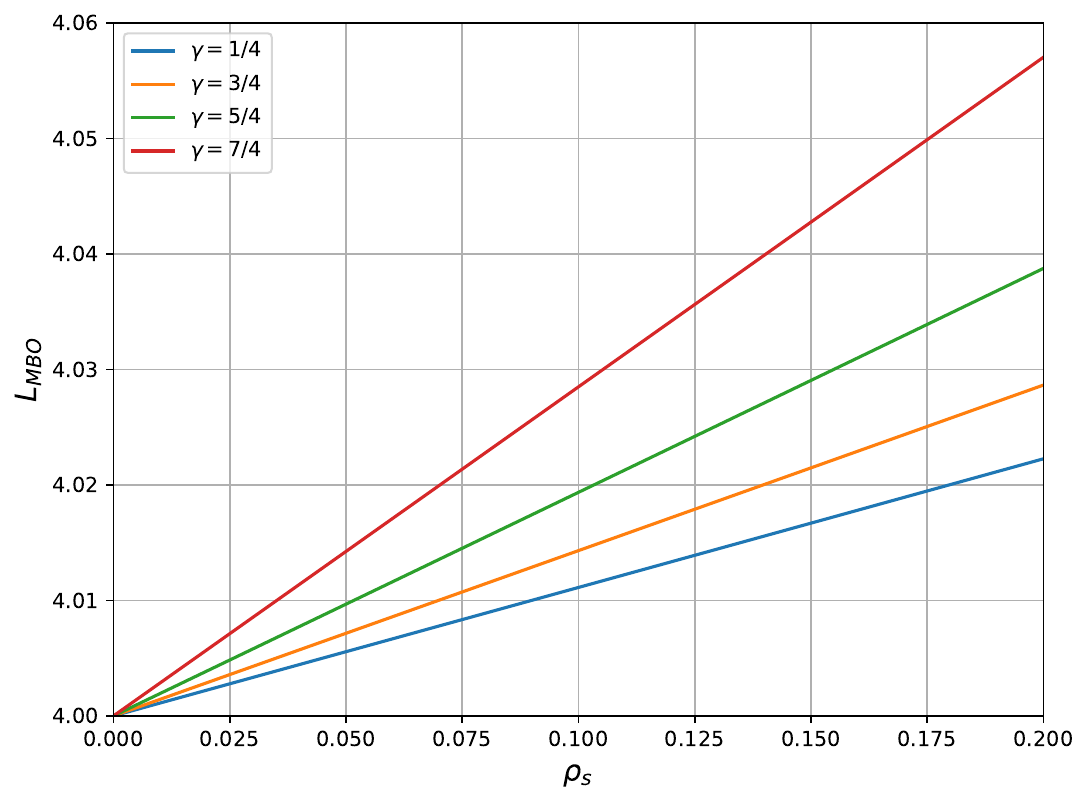}
    \includegraphics[width=0.32\textwidth]{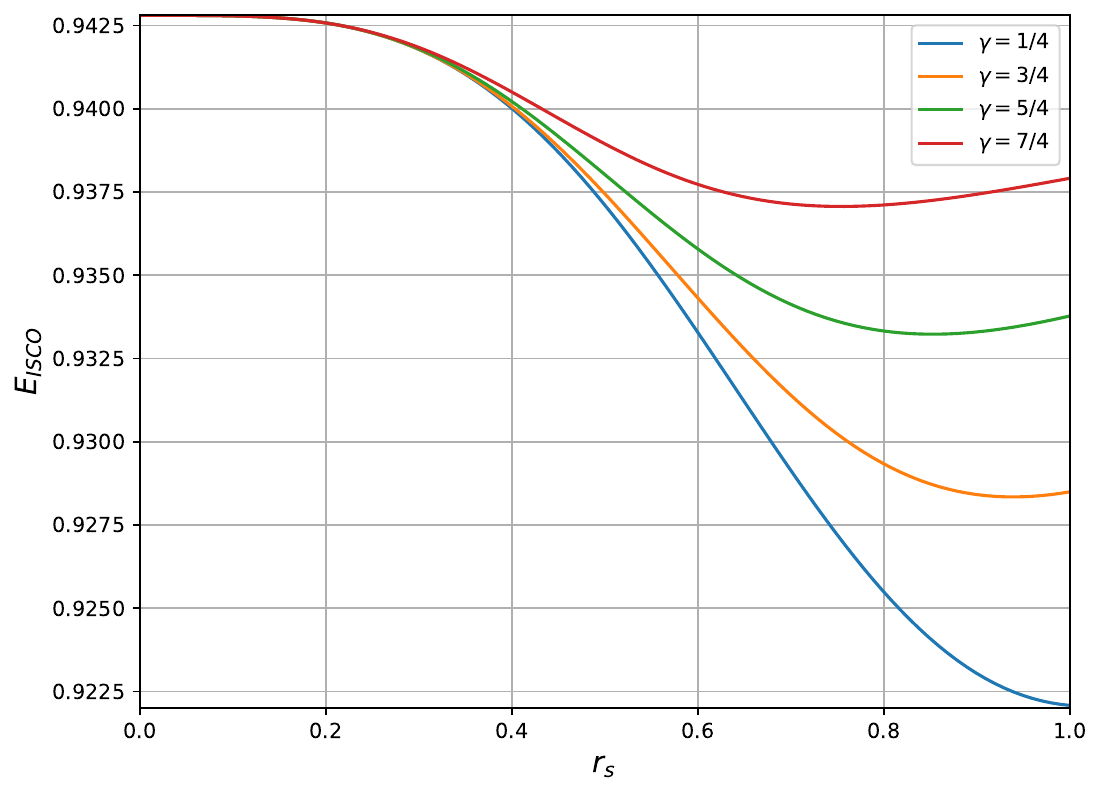}

    \includegraphics[width=0.32\textwidth]{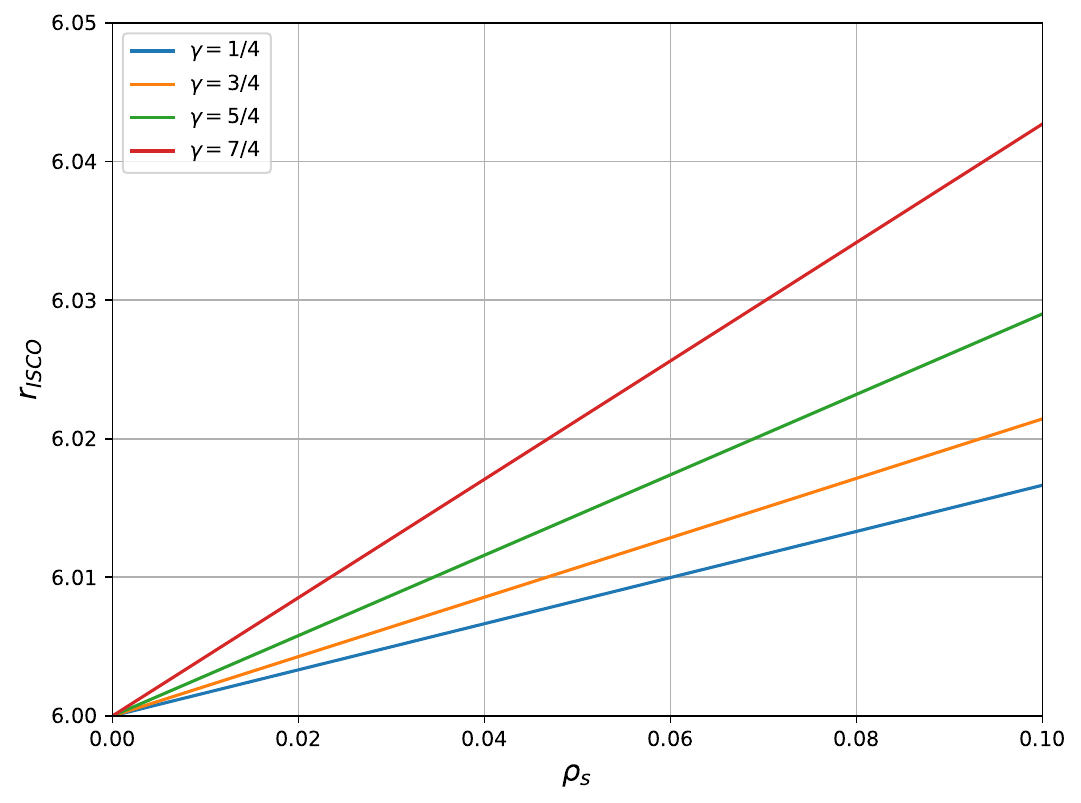}
    \includegraphics[width=0.32\textwidth]{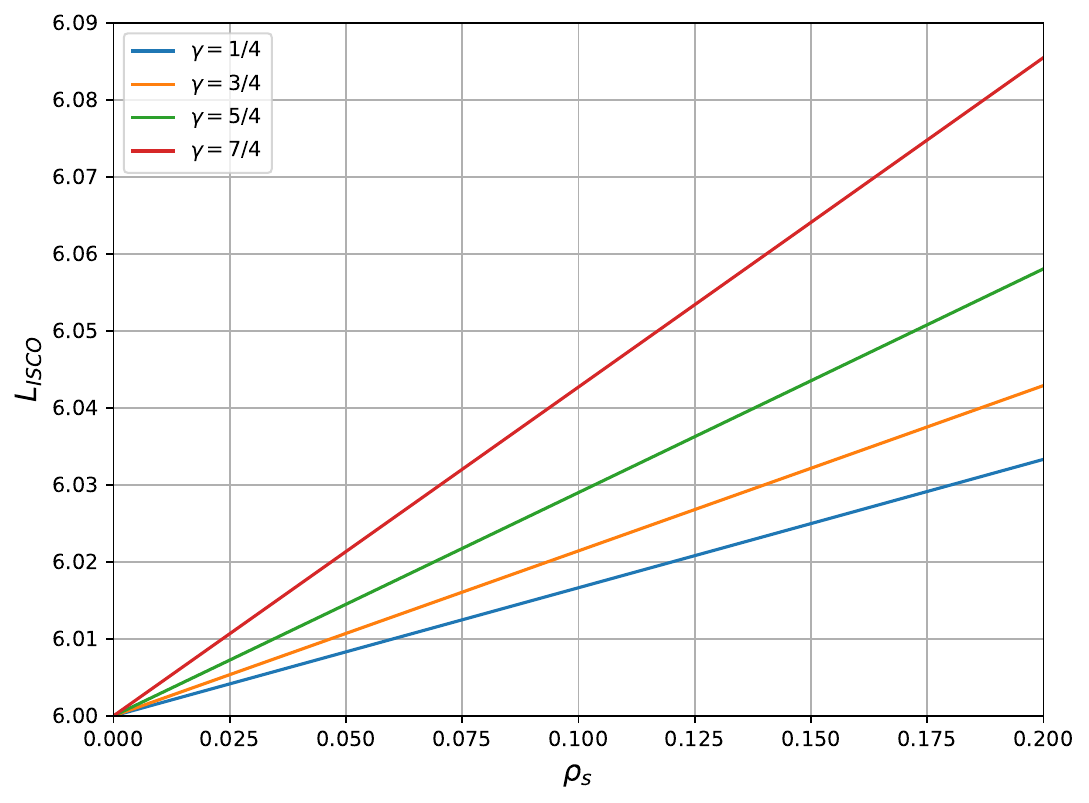}
    \includegraphics[width=0.32\textwidth]{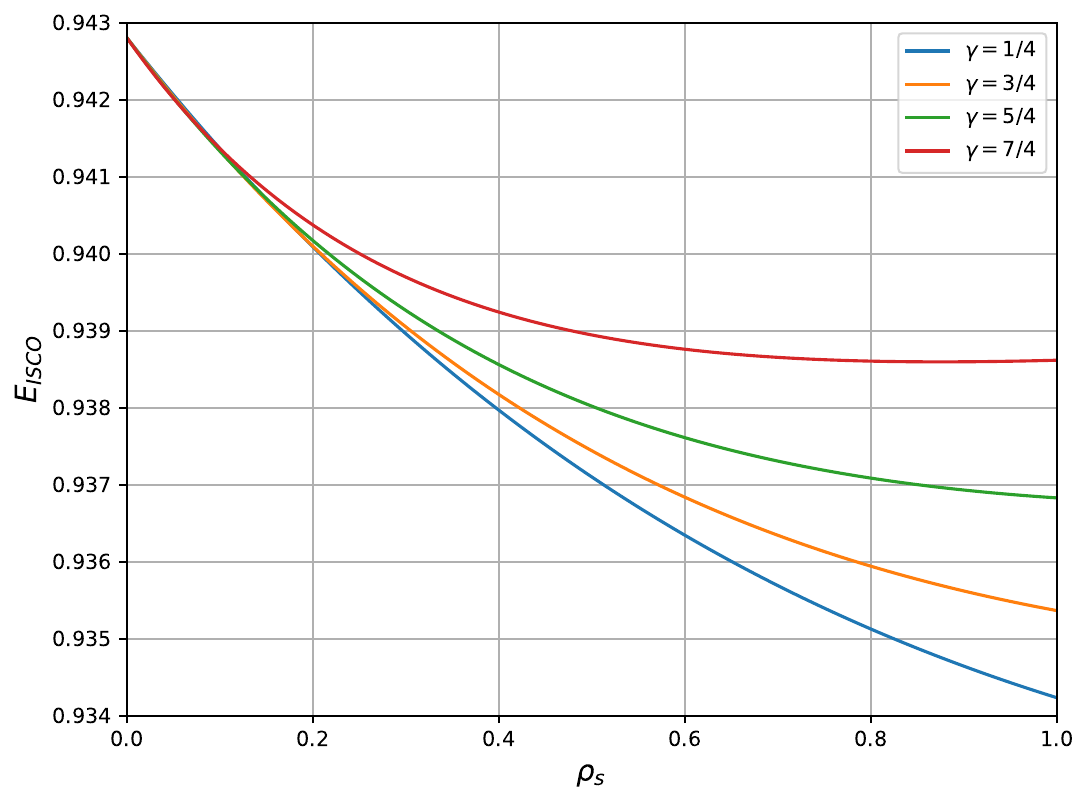}
    \caption{MBO and ISCO parameters for different values of the Dehnen-type halo parameter $\gamma$. The left and middle panels show $r_{\rm MBO}$, $L_{\rm MBO}$, $r_{\rm ISCO}$, and $L_{\rm ISCO}$ as functions of $\rho_s$, while the right panels display $E_{\rm ISCO}$ as a function of $r_s$ (top) and $\rho_s$ (bottom).
    }
\label{fig:isco_mbo}
\end{figure*}
\begin{figure*}[htbp]
\centering
 \includegraphics[scale=0.4]{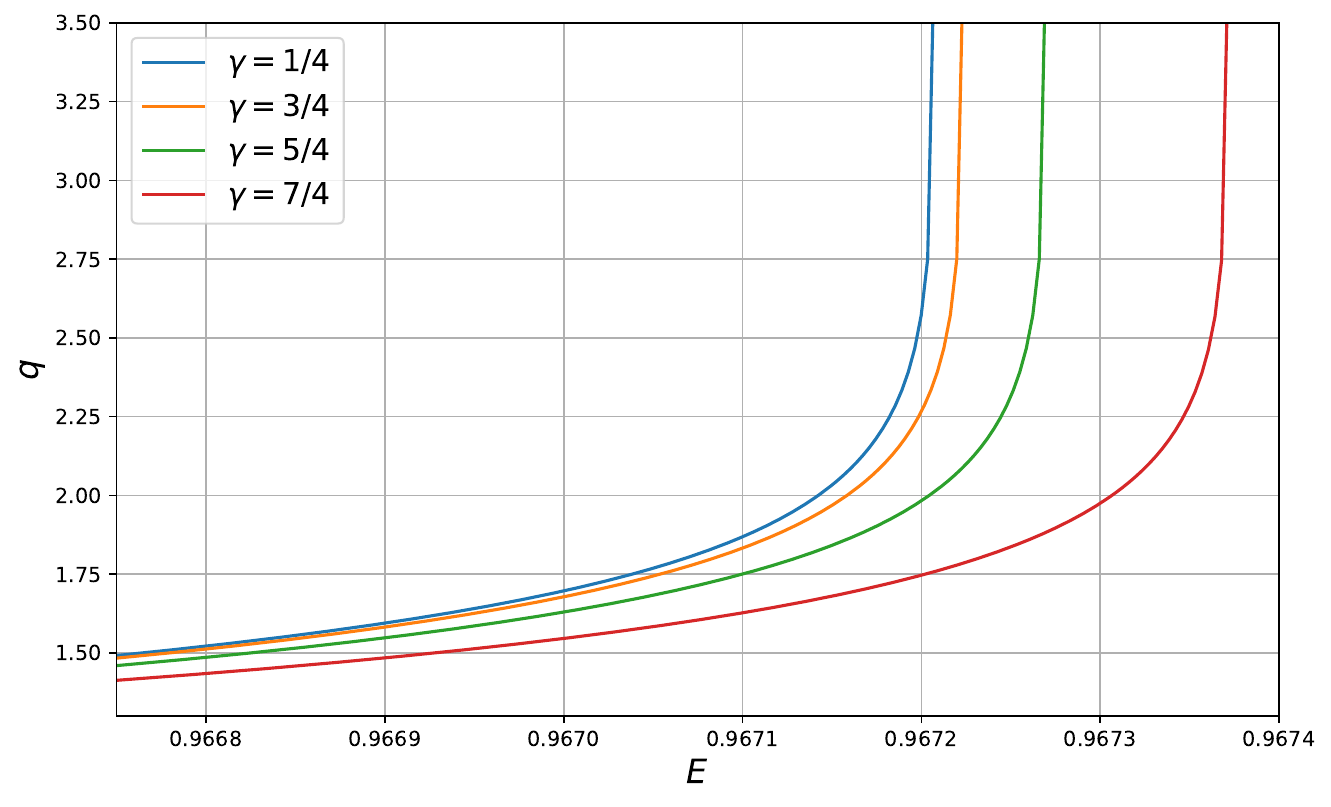}
 \includegraphics[scale=0.4]{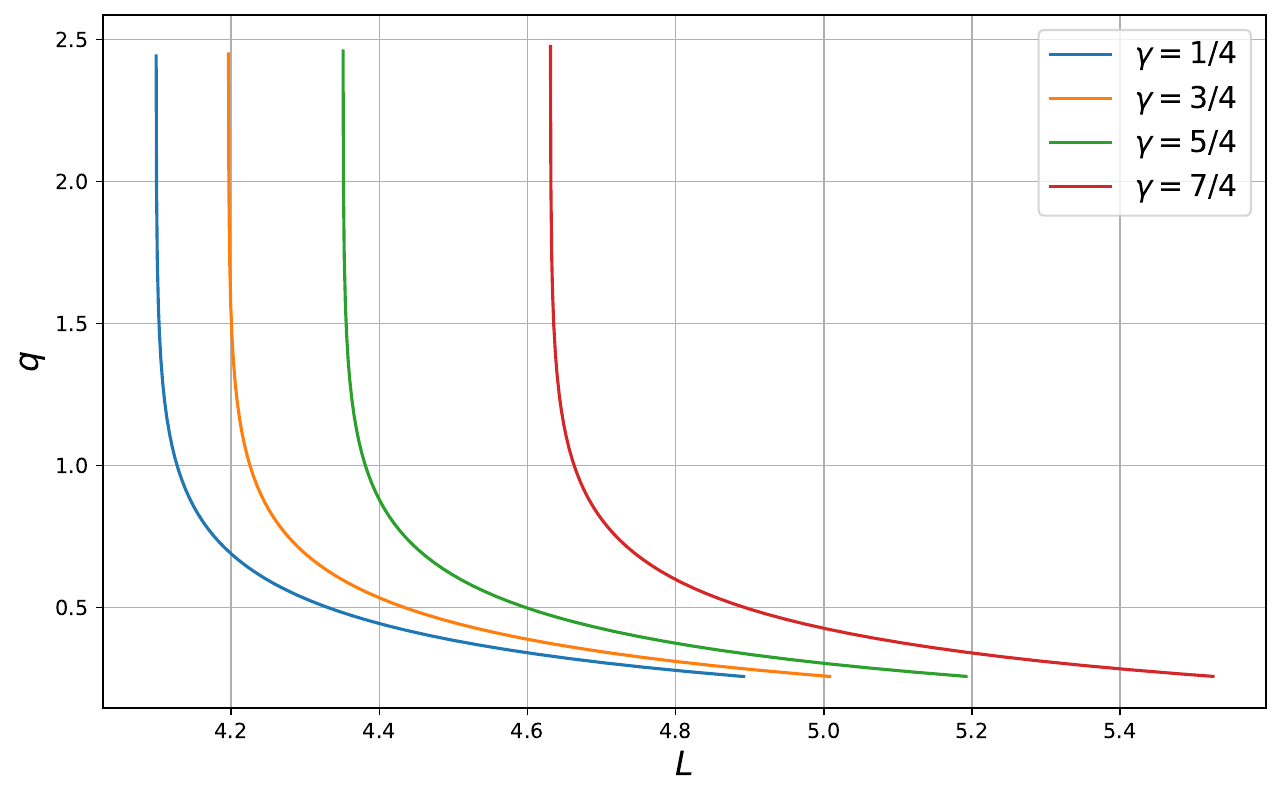}
  
 \caption{Dependence of the rational number $q$ on the energy and angular momentum of periodic orbits for different values of the Dehnen-type DM halo parameter $\gamma$. The left panel shows $q$ as a function of the particle energy $E$, with the angular momentum fixed at $L=(L_{\rm MBO}+L_{\rm ISCO})/2$. The right panel shows $q$ as a function of the orbital angular momentum $L$, with the particle energy fixed at $E=0.96$.}
 \label{fig:q}
\end{figure*}
\renewcommand{\arraystretch}{1.2}
\begin{table*}[t]

\resizebox{1.0\textwidth}{!}{
\begin{tabular}{|c|c|c|c|c|c|c|c|c|c|}
\hline
$\gamma$ & $L$ & $E_{(1,1,0)}$ & $E_{(1,2,0)}$ & $E_{(2,1,1)}$ & $E_{(2,2,1)}$ & $E_{(3,1,2)}$ & $E_{(3,2,2)}$ & $E_{(4,1,3)}$ & $E_{(4,2,3)}$ \\ \hline

$1/4$ & 4.362009 & 0.962528 & 0.965791 & 0.965390 & 0.965850 & 0.965612 & 0.965855 & 0.965680 & 0.965857 \\ \hline

$3/4$ & 4.554315 & 0.962627 & 0.965901 & 0.965498 & 0.965960 & 0.965721 & 0.965965 & 0.965789 & 0.965967 \\ \hline

$5/4$ & 4.859526 & 0.962842 & 0.966109 & 0.965707 & 0.966169 & 0.965930 & 0.966174 & 0.965998 & 0.966175 \\ \hline

$7/4$ & 5.411331 & 0.963242 & 0.966474 & 0.966077 & 0.966533 & 0.966297 & 0.966538 & 0.966364 & 0.966539 \\ \hline

\end{tabular}
}
\centering
\caption{The values of the energy $E$ for different periodic orbits characterized by $(z,w,v)$. The parameter $\gamma$ varies as shown. Here, $L$ is determined based on the orbital parameters.}
\label{table1}
\end{table*}
\renewcommand{\arraystretch}{1.2}
\begin{table*}[t]

\resizebox{1.0\textwidth}{!}{
\begin{tabular}{|c|c|c|c|c|c|c|c|c|c|}
\hline
$\gamma$ & $E$ & $L_{(1,1,0)}$ & $L_{(1,2,0)}$ & $L_{(2,1,1)}$ & $L_{(2,2,1)}$ & $L_{(3,1,2)}$ & $L_{(3,2,2)}$ & $L_{(4,1,3)}$ & $L_{(4,2,3)}$ \\ \hline

$1/4$ & 0.968986 & 4.425121 & 4.394430 & 4.398199 & 4.393888 & 4.396101 & 4.393845 & 4.395465 & 4.393831 \\ \hline

$3/4$ & 0.969088 & 4.620258 & 4.588102 & 4.592057 & 4.587532 & 4.589856 & 4.587486 & 4.589188 & 4.587472 \\ \hline

$5/4$ & 0.969282 & 4.929814 & 4.895499 & 4.899721 & 4.894890 & 4.897372 & 4.894841 & 4.896659 & 4.894827 \\ \hline

$7/4$ & 0.969622 & 5.489270 & 5.451285 & 5.455950 & 5.450614 & 5.453354 & 5.450560 & 5.452566 & 5.450544 \\ \hline

\end{tabular}
}
\centering
\caption{The values of the angular momentum $L$ for different periodic orbits characterized by $(z,w,v)$, while the energy $E$ is fixed for each value of $\gamma$. Here $E=(E_{MBO}+E_{ISCO})/2$}
\label{table2}
\end{table*}

We now focus on periodic orbits around a Schwarzschild BH embedded in a Dehnen-type $(1,4,\gamma)$ DM halo. Since periodic orbits are a special class of bound motion, the conserved energy $E$ and orbital angular momentum $L$ of the test particle must satisfy \cite{Dadhich22a}
\begin{equation}
L \geq L_{\rm ISCO} 
\qquad\mbox{and} \qquad
E_{\rm ISCO} \leq E \leq E_{\rm MBO}=1\, ,
\end{equation}
where $E_{\rm ISCO}$ and $L_{\rm ISCO}$ are the particle energy and angular momentum at the ISCO, respectively, and $E_{\rm MBO}$ is the particle energy at the MBO.
The marginally bound orbit (MBO) and the innermost stable circular orbit (ISCO) are determined from the effective potential. For the MBO, the following conditions must be satisfied:
\begin{equation}
V_{\rm eff}(r_{\rm MBO})=1,
\qquad
\frac{dV_{\rm eff}}{dr}\bigg|_{r=r_{\rm MBO}}=0 .
\end{equation}
The ISCO is obtained from the following conditions
\begin{equation}
\frac{dV_{\rm eff}}{dr}\bigg|_{r=r_{\rm ISCO}}=0,
\qquad
\frac{d^2V_{\rm eff}}{dr^2}\bigg|_{r=r_{\rm ISCO}}=0 .
\end{equation}
Using the above conditions, we numerically investigate the effects of the DM halo on the ISCO and MBO parameters. The dependence of the MBO and ISCO parameters on the halo characteristic density $\rho_s$, scale radius $r_s$, and halo profile parameter $\gamma$ is presented in Fig.~\ref{fig:isco_mbo}. As shown in Fig.~\ref{fig:isco_mbo}, the radii and angular momenta associated with the MBO and ISCO increase with the DM halo density $\rho_s$. In particular, $r_{\rm MBO}$, $L_{\rm MBO}$, $r_{\rm ISCO}$, and $L_{\rm ISCO}$ increase as $\rho_s$ increases. For fixed $\rho_s$, larger values of the Dehnen-type halo parameter $\gamma$ produce larger values of these quantities, indicating that the MBO and ISCO are shifted outward and that particles require greater angular momentum to remain on these orbits.
The behavior of the ISCO energy is shown in the right panels. The energy $E_{\rm ISCO}$ decreases with increasing scale radius $r_s$ and halo density $\rho_s$. However, larger values of $\gamma$ lead to relatively higher values of $E_{\rm ISCO}$ over the considered parameter range. This suggests that the Dehnen-type DM halo changes not only the positions of the MBO and ISCO but also the binding energy of particles at the ISCO.
\begin{figure*}[htbp]
    \centering
    \includegraphics[width=0.32\textwidth]{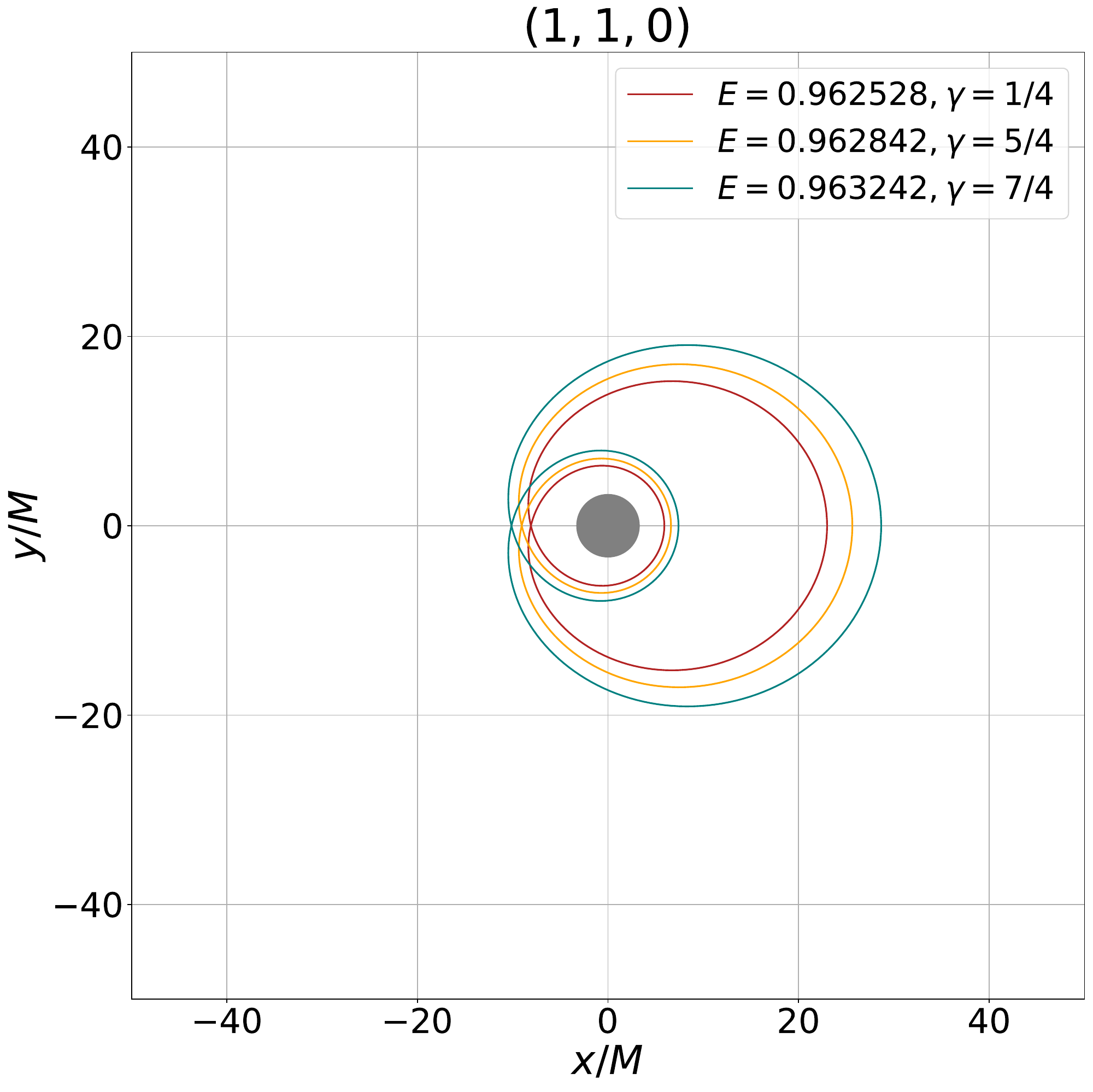} \hfill
    \includegraphics[width=0.32\textwidth]{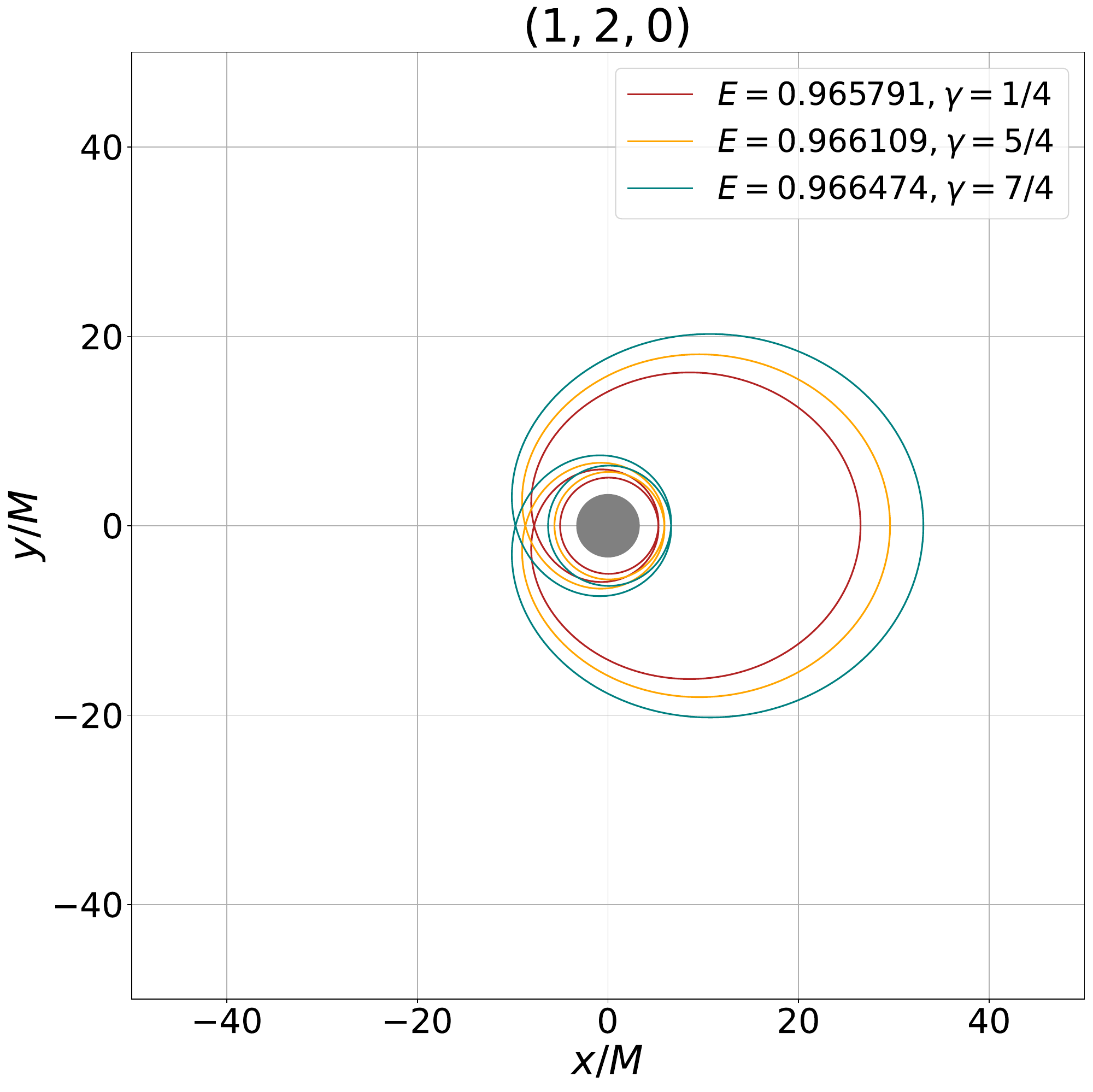} \hfill
    \includegraphics[width=0.32\textwidth]{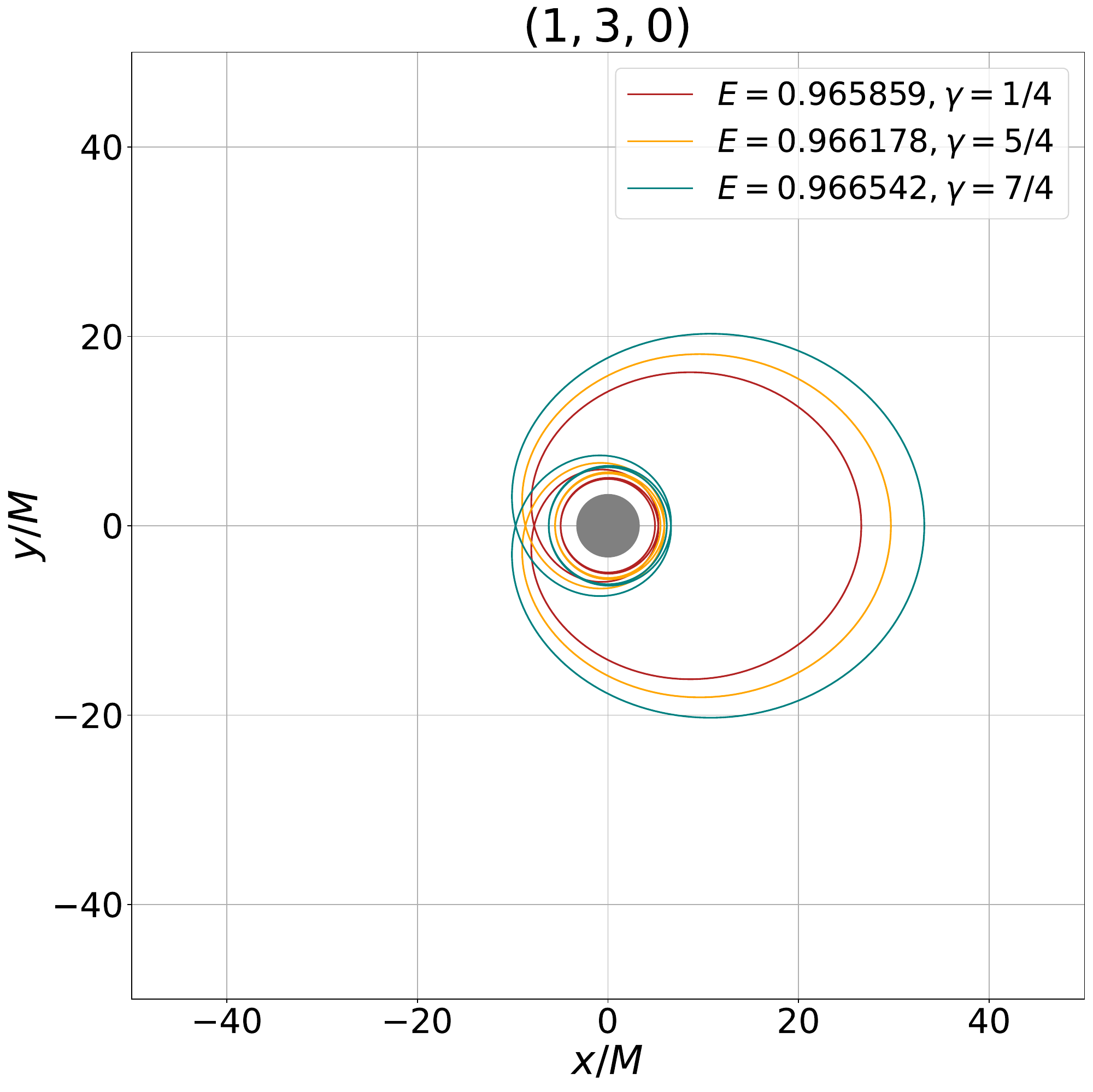} \\
    
    \vspace{0.2cm} 
    \includegraphics[width=0.32\textwidth]{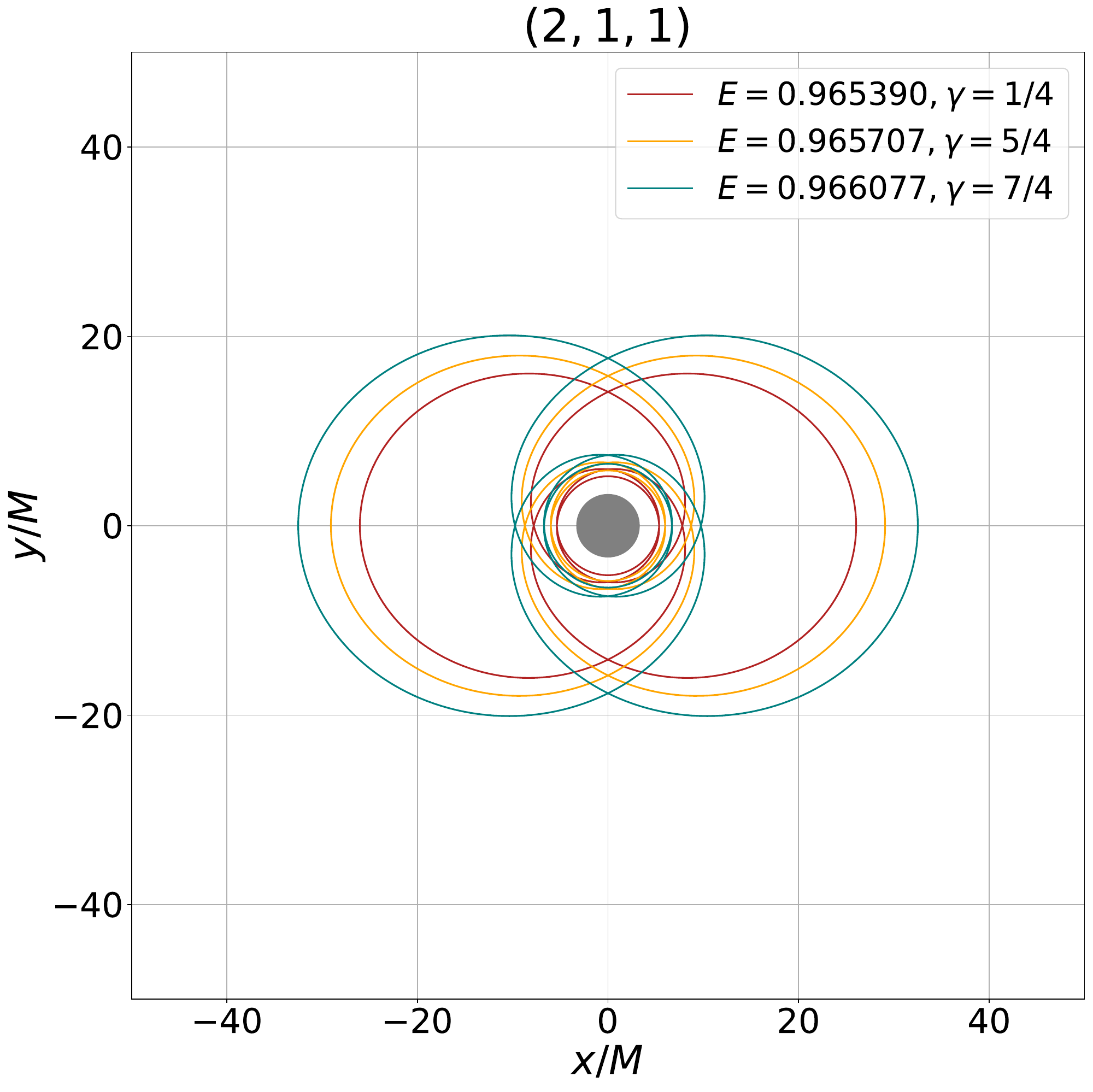} \hfill
    \includegraphics[width=0.32\textwidth]{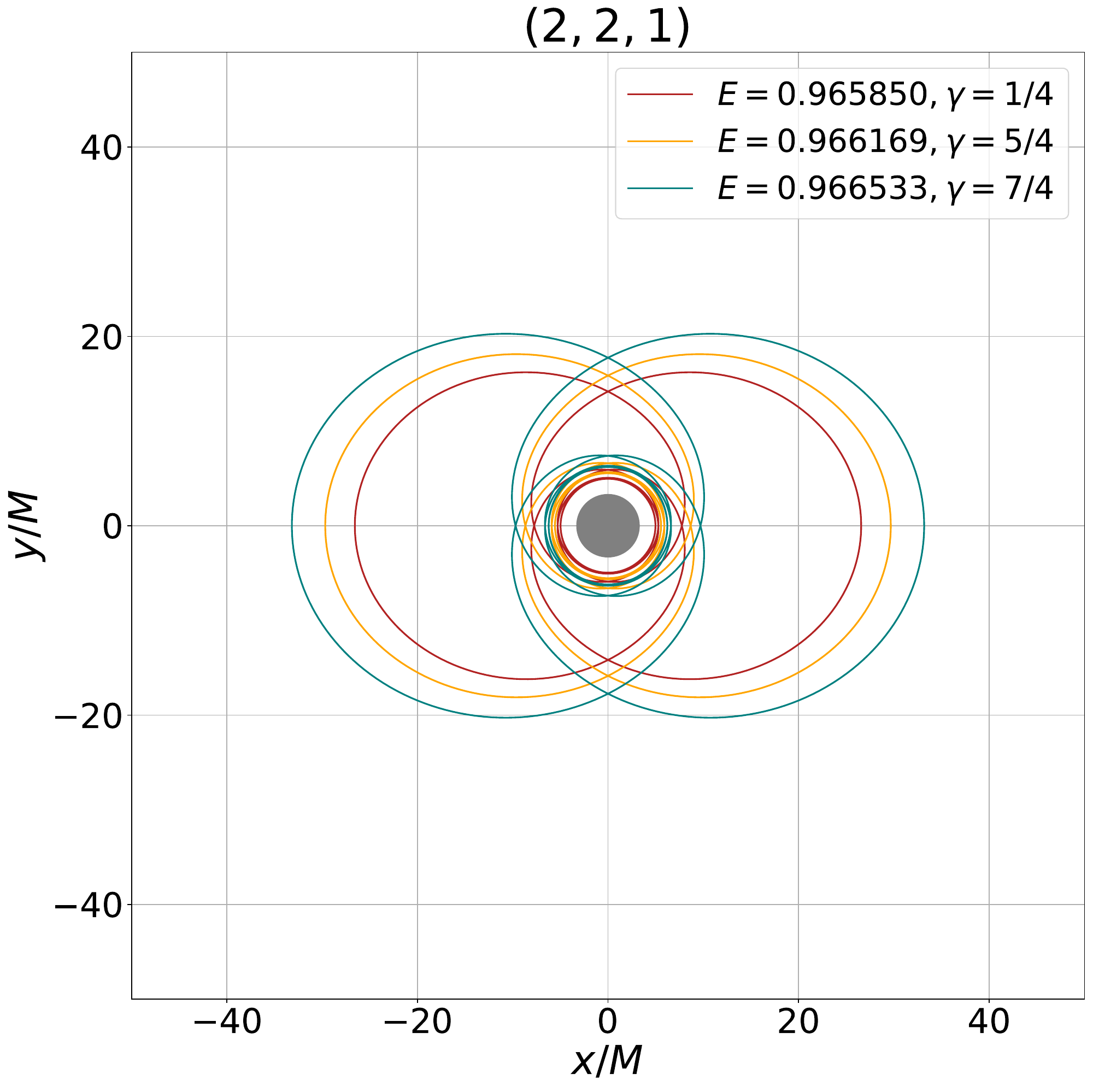} \hfill
    \includegraphics[width=0.32\textwidth]{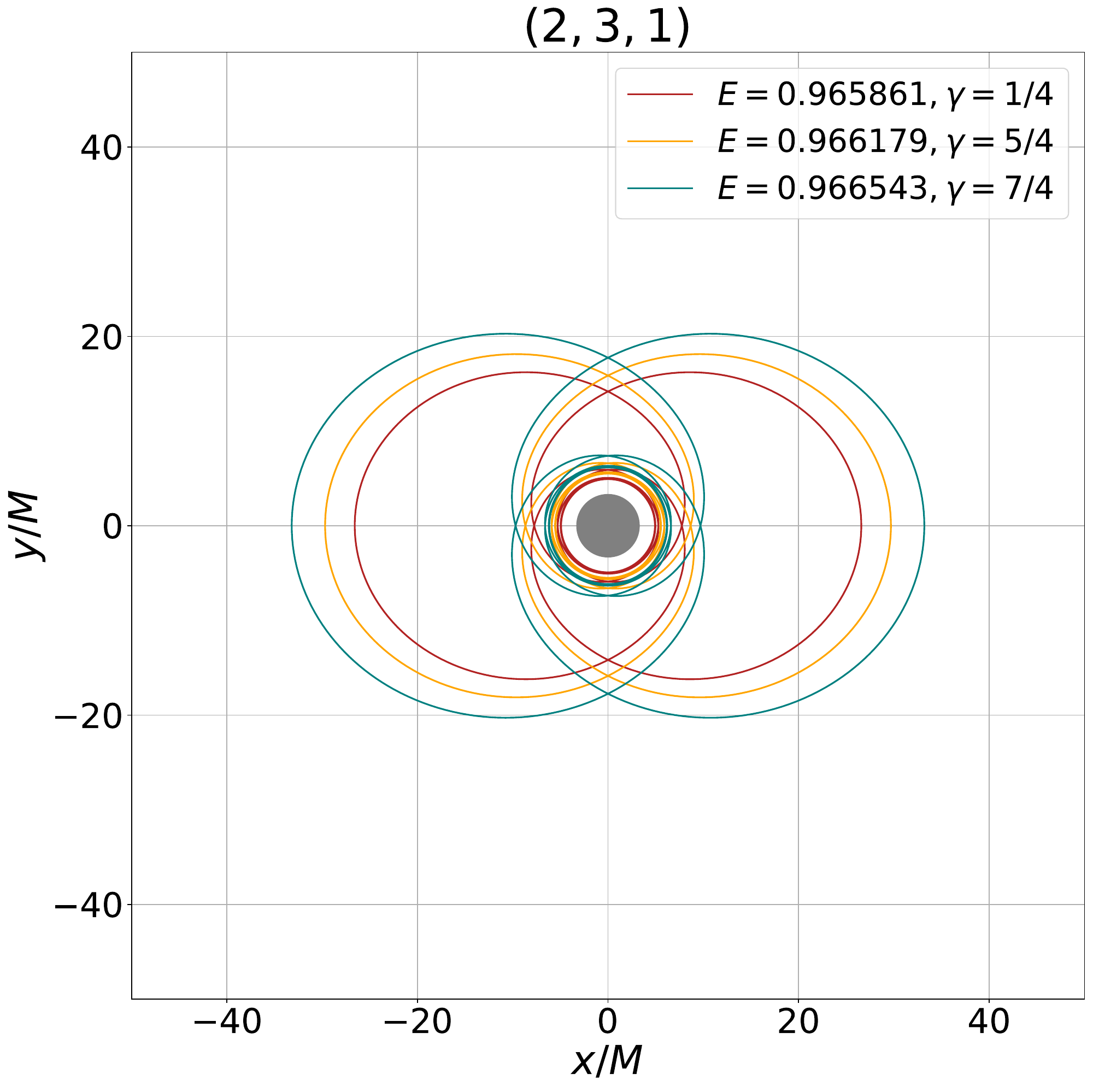} \\
    
    \vspace{0.2cm} 
    \includegraphics[width=0.32\textwidth]{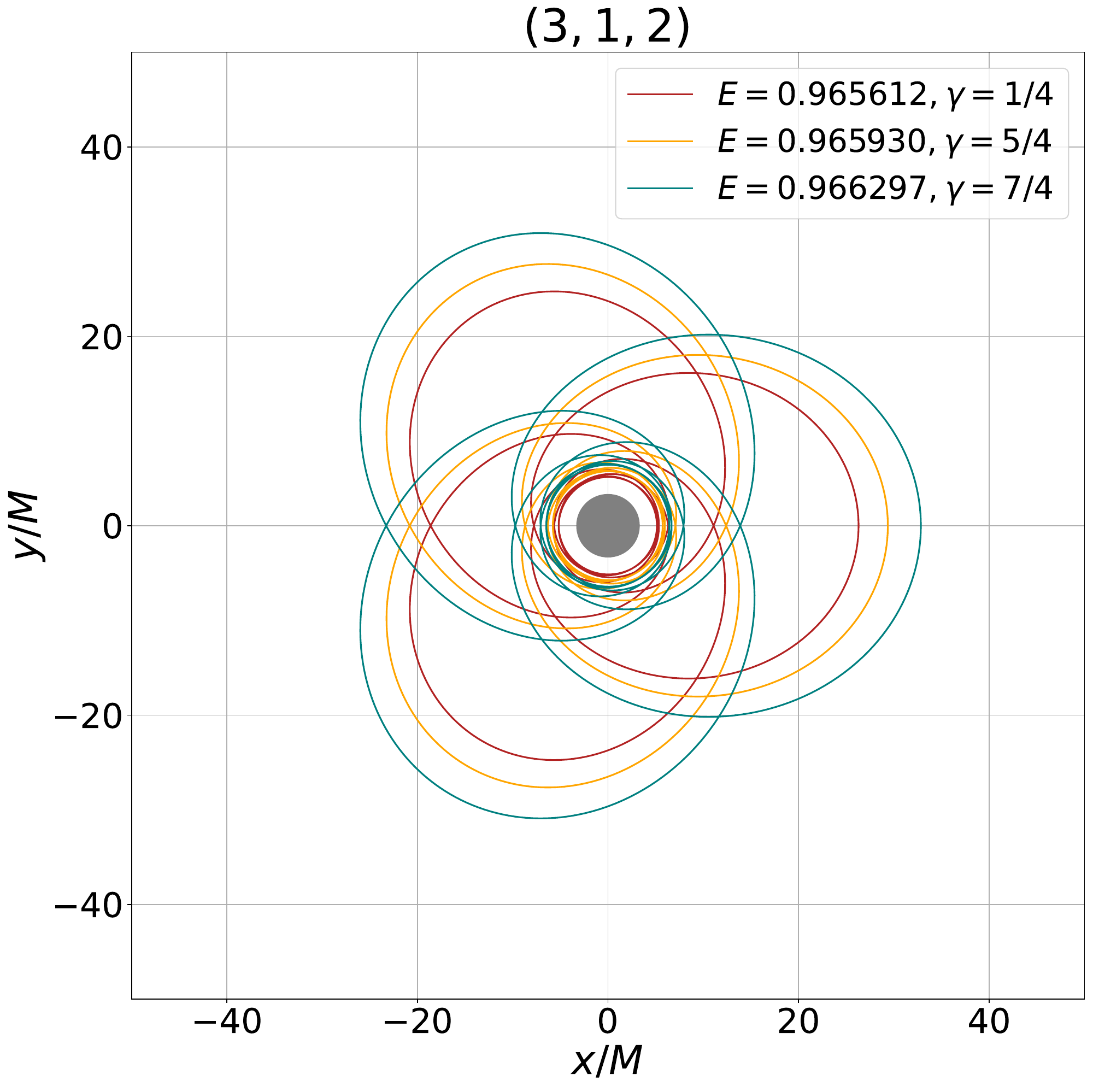} \hfill
    \includegraphics[width=0.32\textwidth]{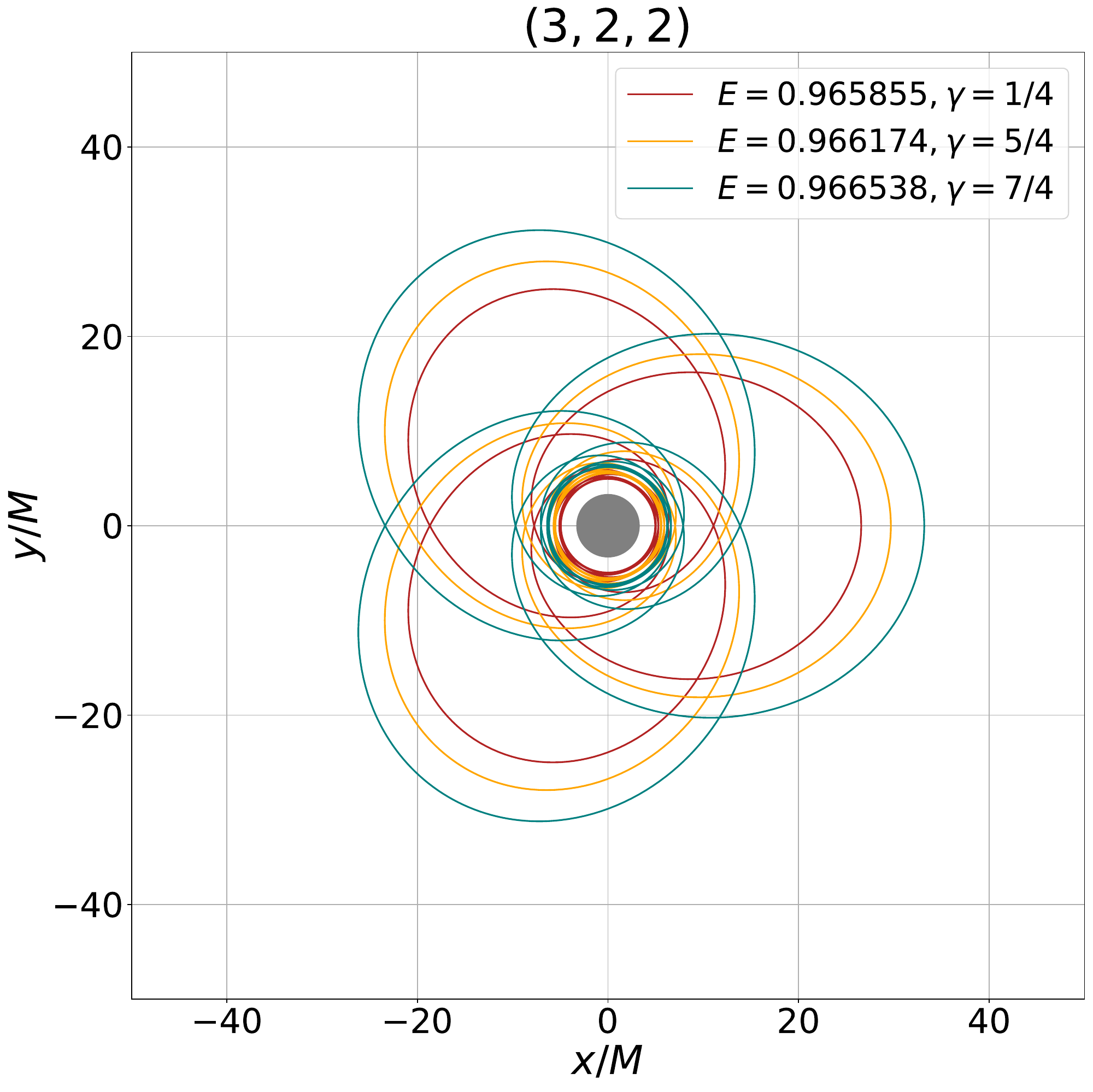} \hfill
    \includegraphics[width=0.32\textwidth]{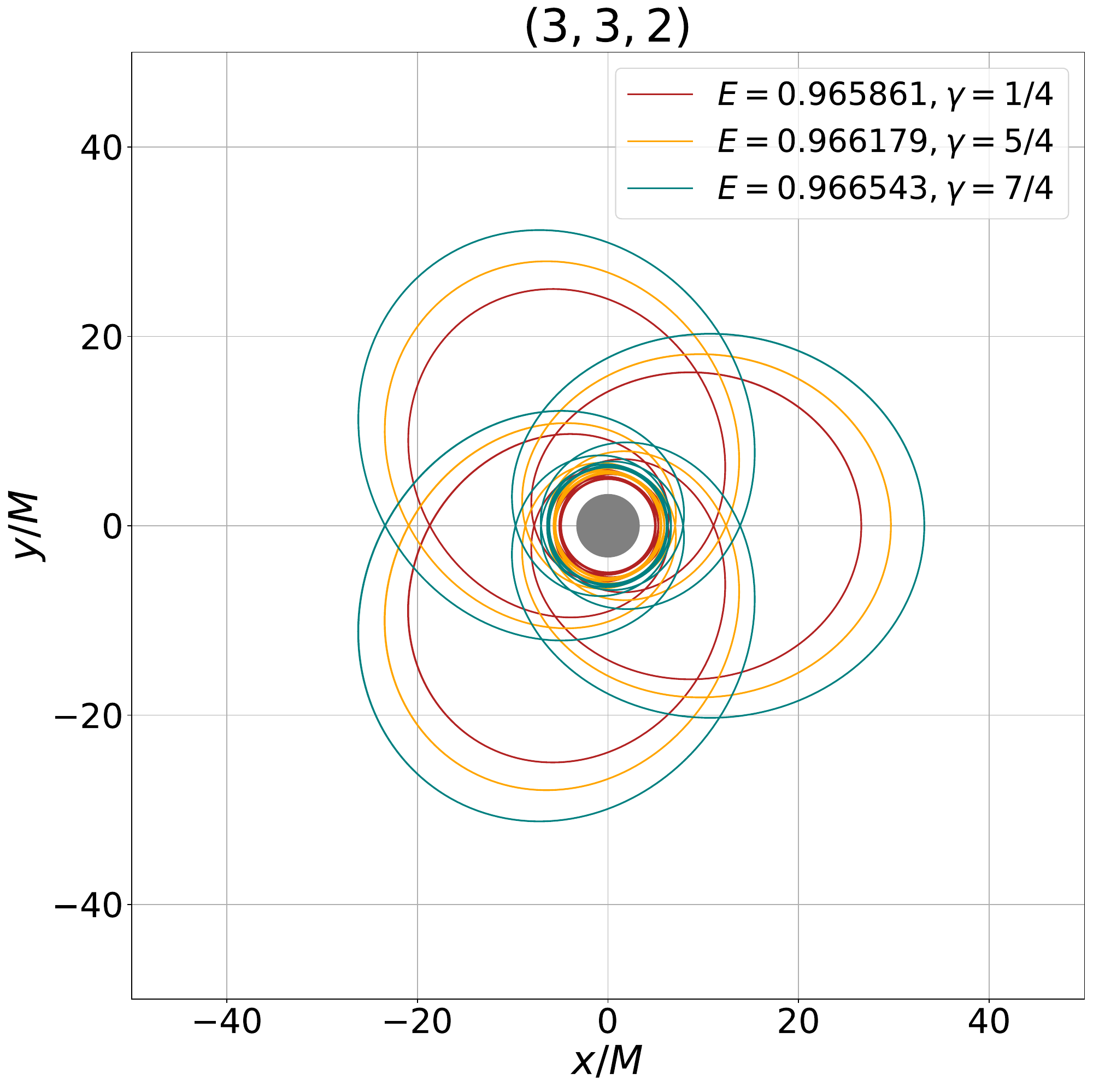} \\
    
    \vspace{0.2cm} 
    \includegraphics[width=0.32\textwidth]{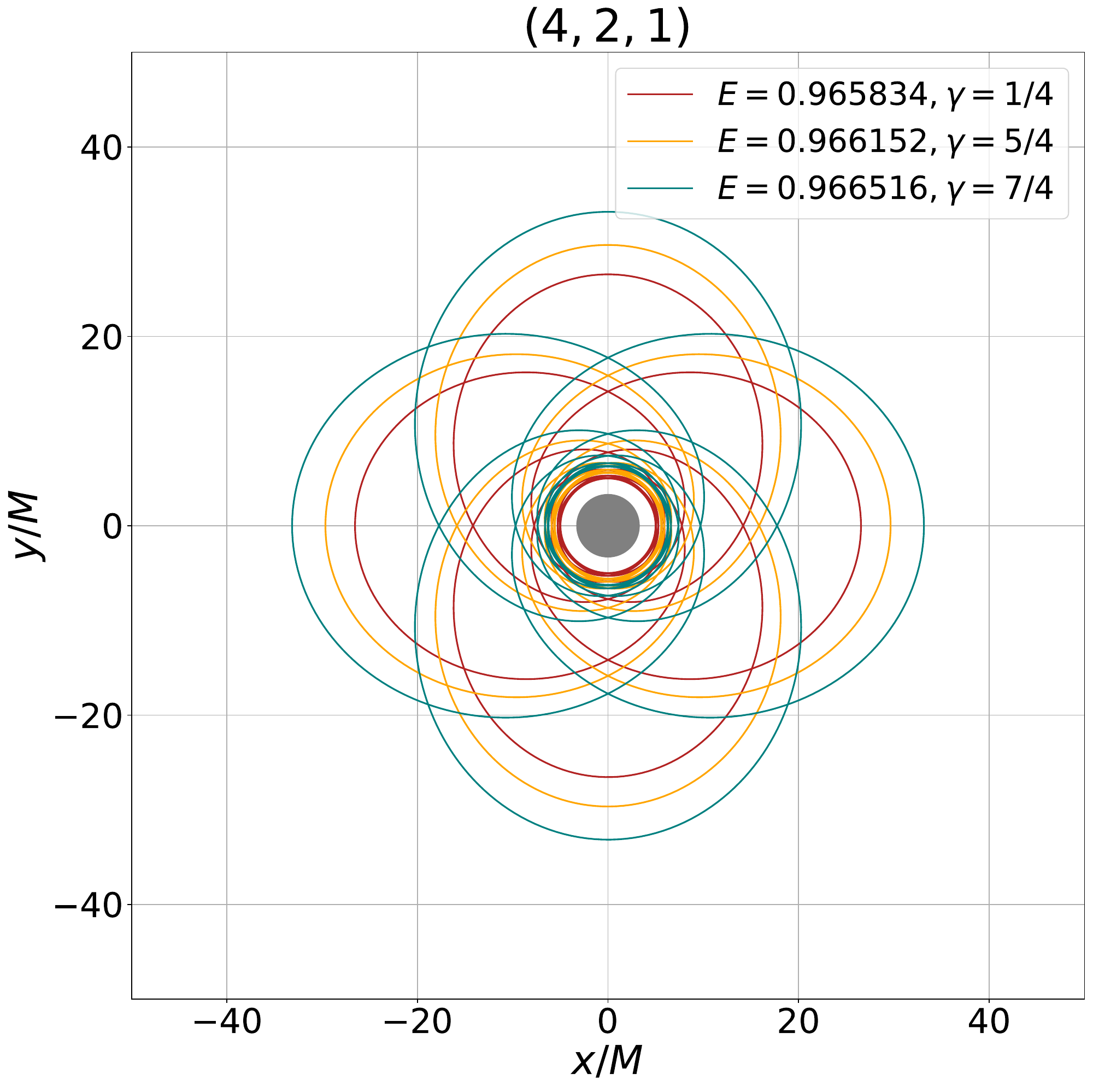} \hfill
    \includegraphics[width=0.32\textwidth]{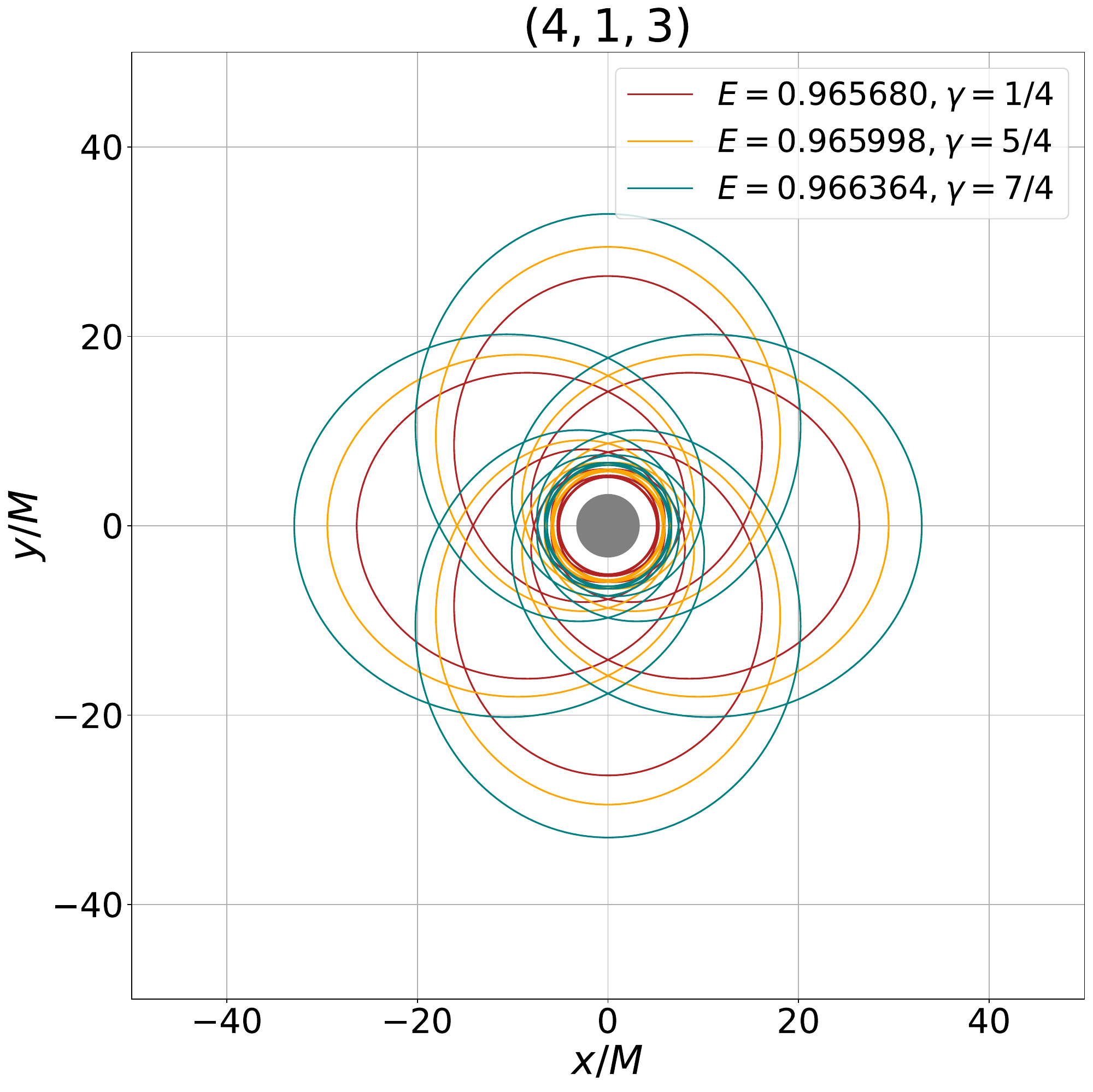} \hfill
    \includegraphics[width=0.32\textwidth]{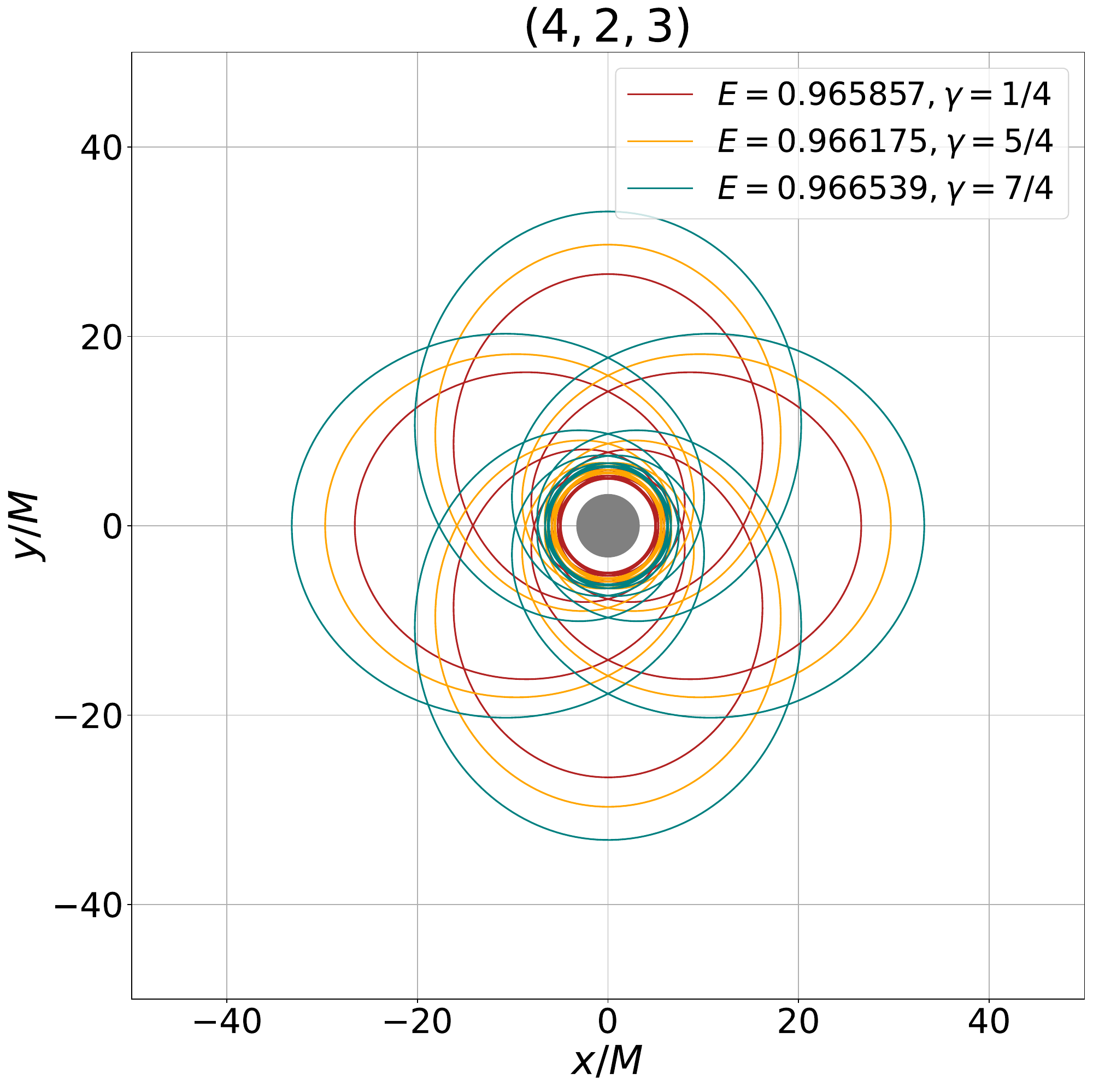}
    
    \caption{Periodic orbits around a generalized Schwarzschild-like BH surrounded by a Dehnen-type DM halo are shown for various combinations of $(z,w,v)$ and $\gamma$. In all cases, we set $\rho_s=0.4$, $r_s=0.5$, and $L=(L_{\rm MBO}+L_{\rm ISCO})/2$.}
    \label{fig:periodic}
\end{figure*}
\begin{figure*}[htbp]
    \centering
    \includegraphics[width=0.32\textwidth]{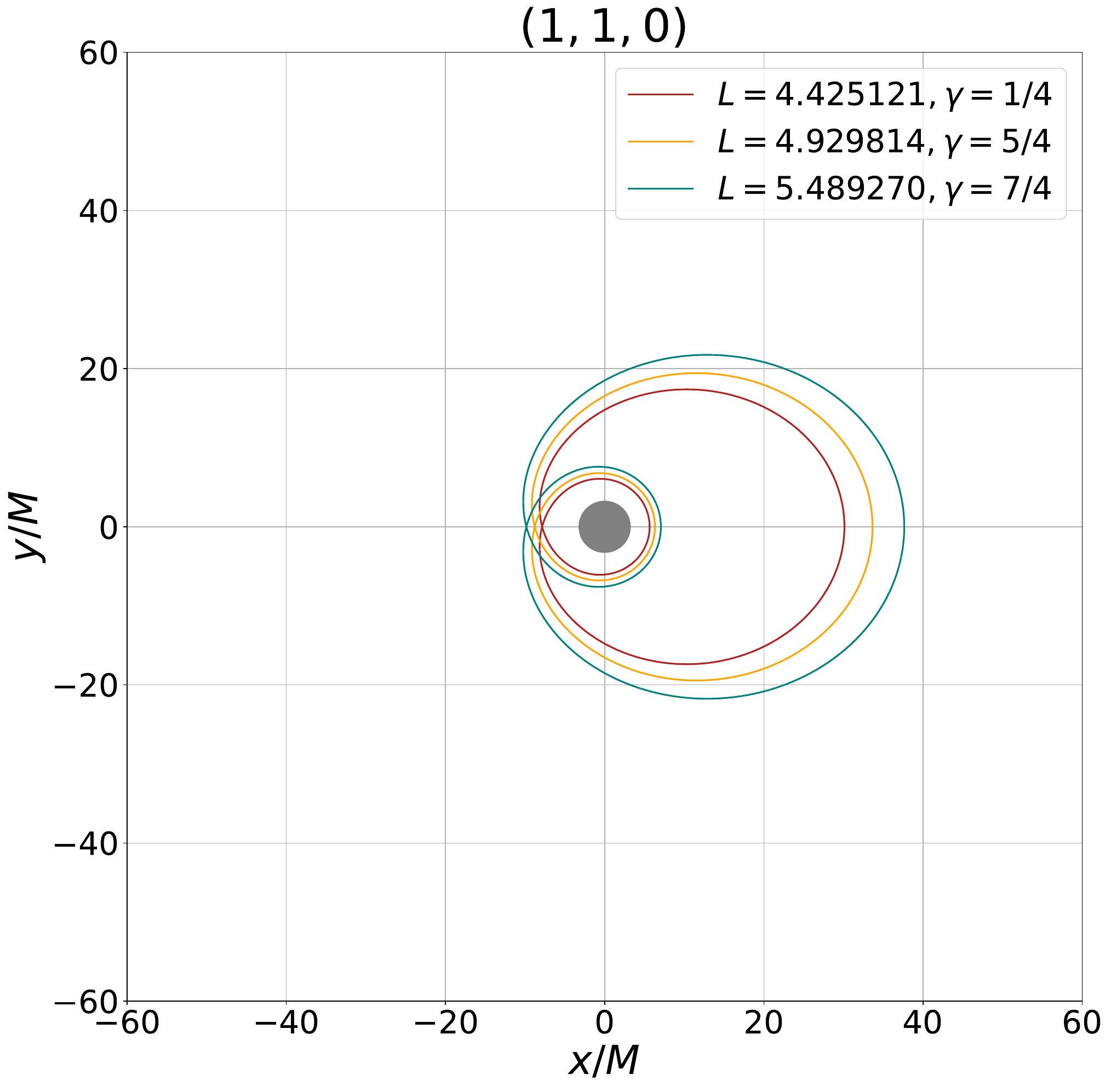} \hfill
    \includegraphics[width=0.32\textwidth]{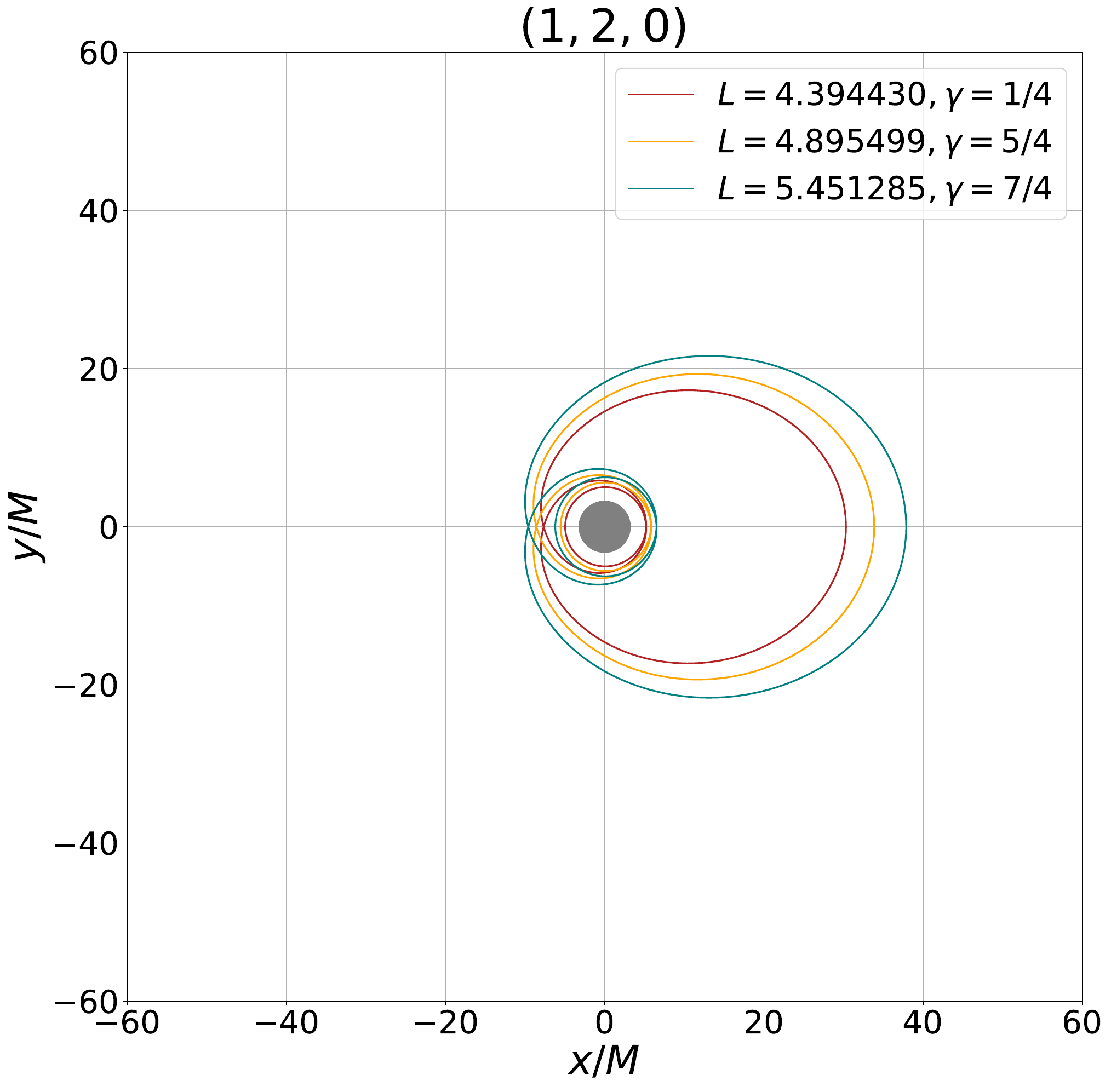} \hfill
    \includegraphics[width=0.32\textwidth]{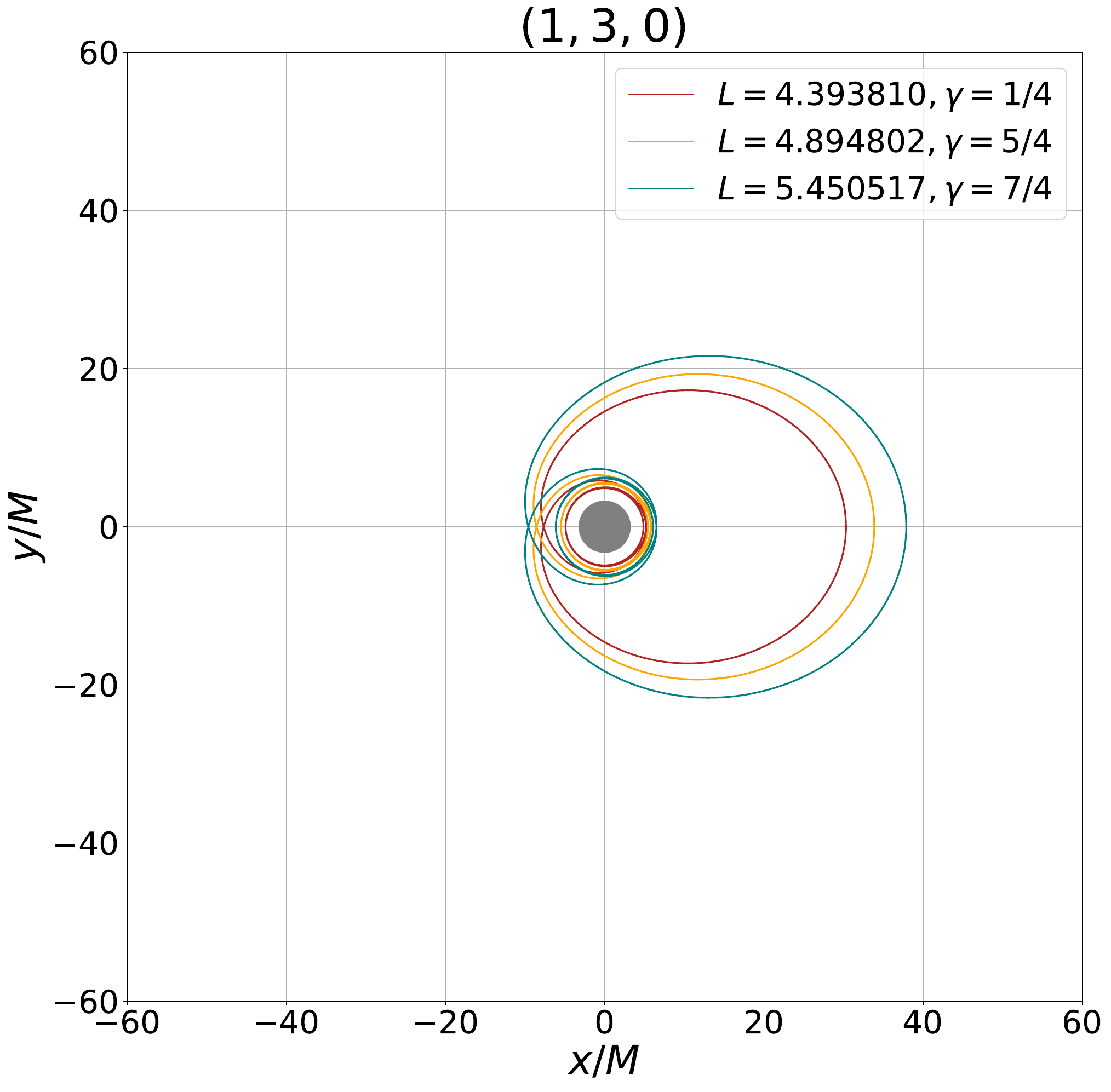} \\
    
    \vspace{0.2cm} 
    \includegraphics[width=0.32\textwidth]{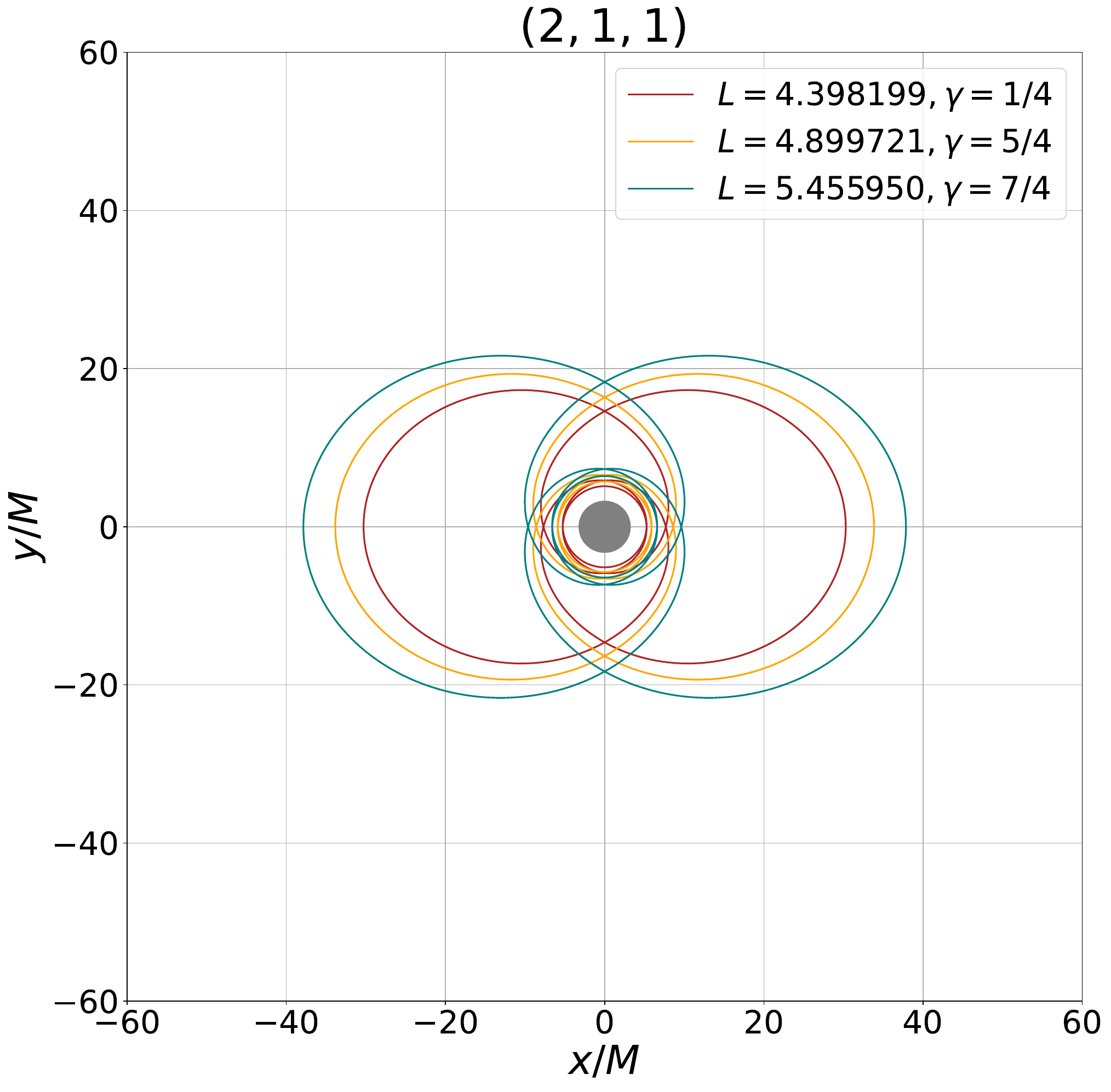} \hfill
    \includegraphics[width=0.32\textwidth]{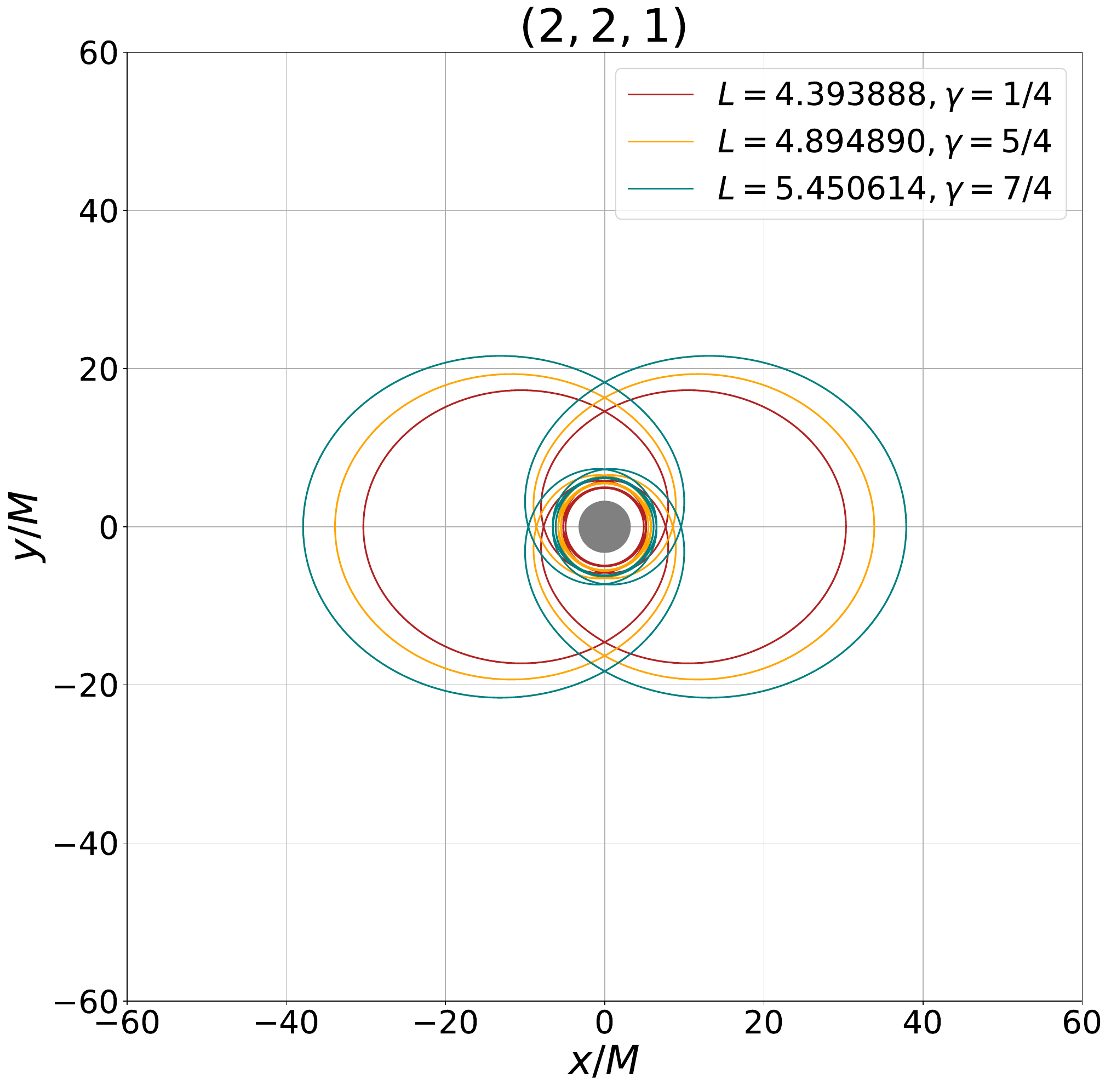} \hfill
    \includegraphics[width=0.32\textwidth]{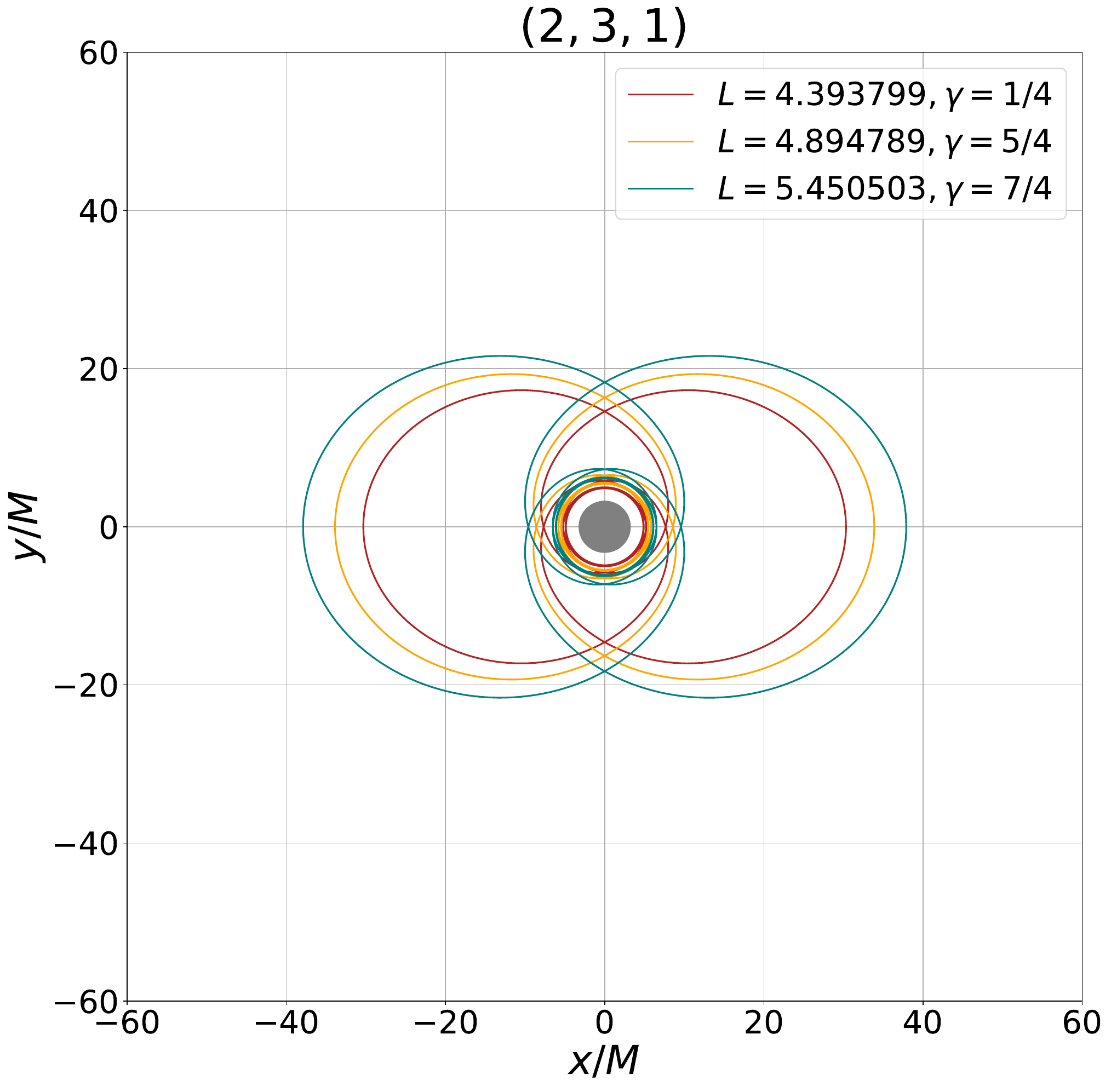} \\
    
    \vspace{0.2cm} 
    \includegraphics[width=0.32\textwidth]{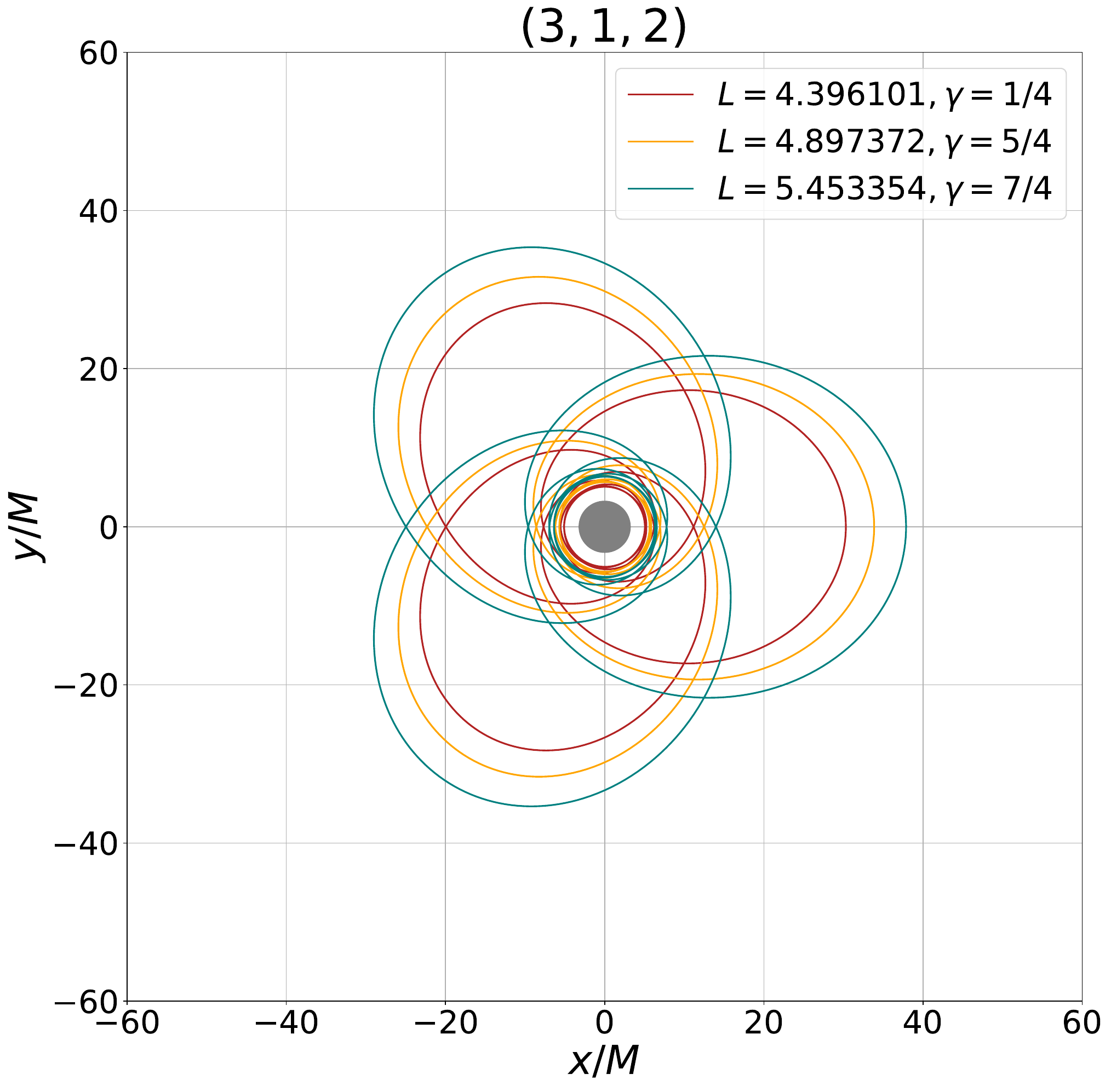} \hfill
    \includegraphics[width=0.32\textwidth]{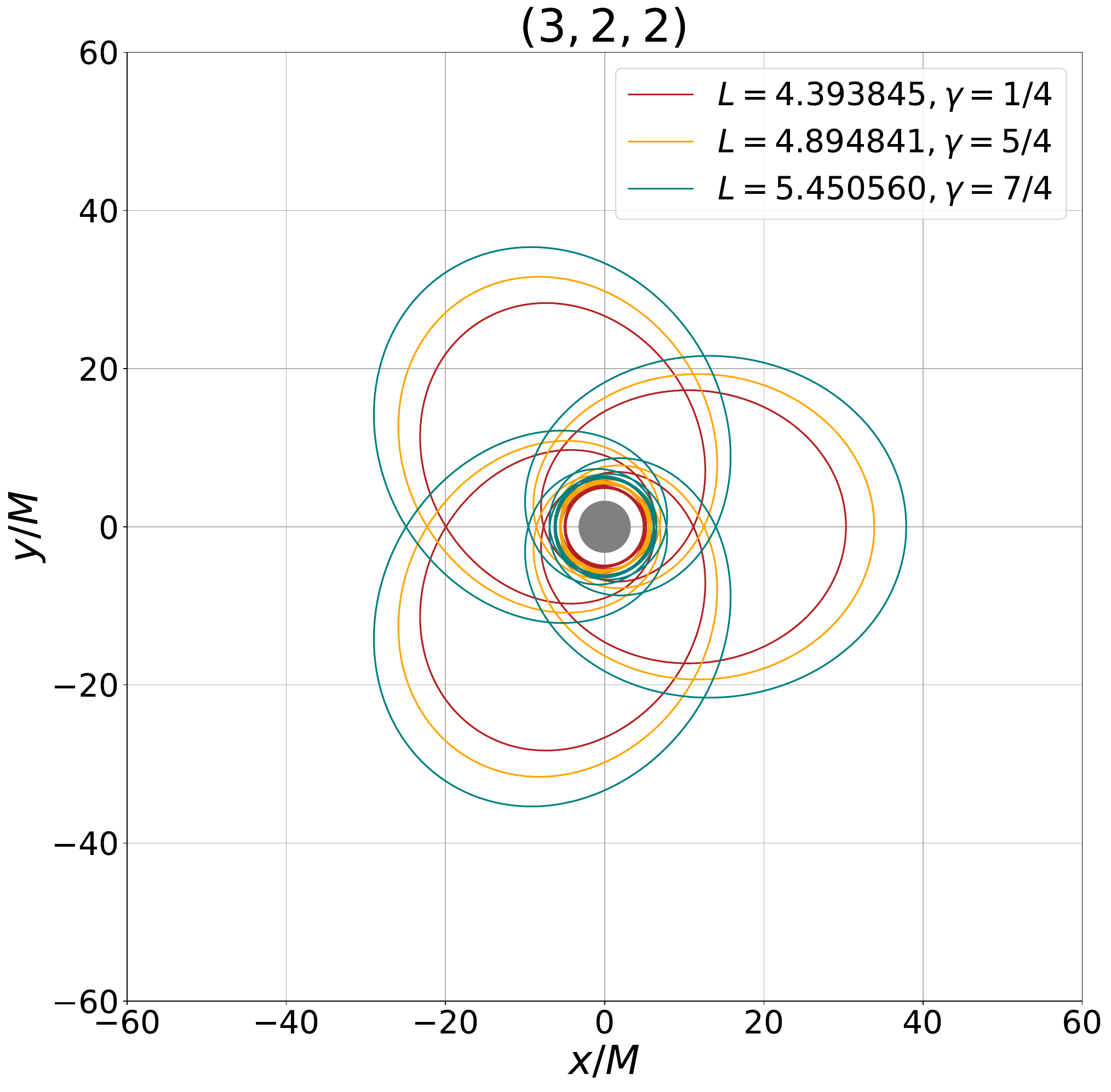} \hfill
    \includegraphics[width=0.32\textwidth]{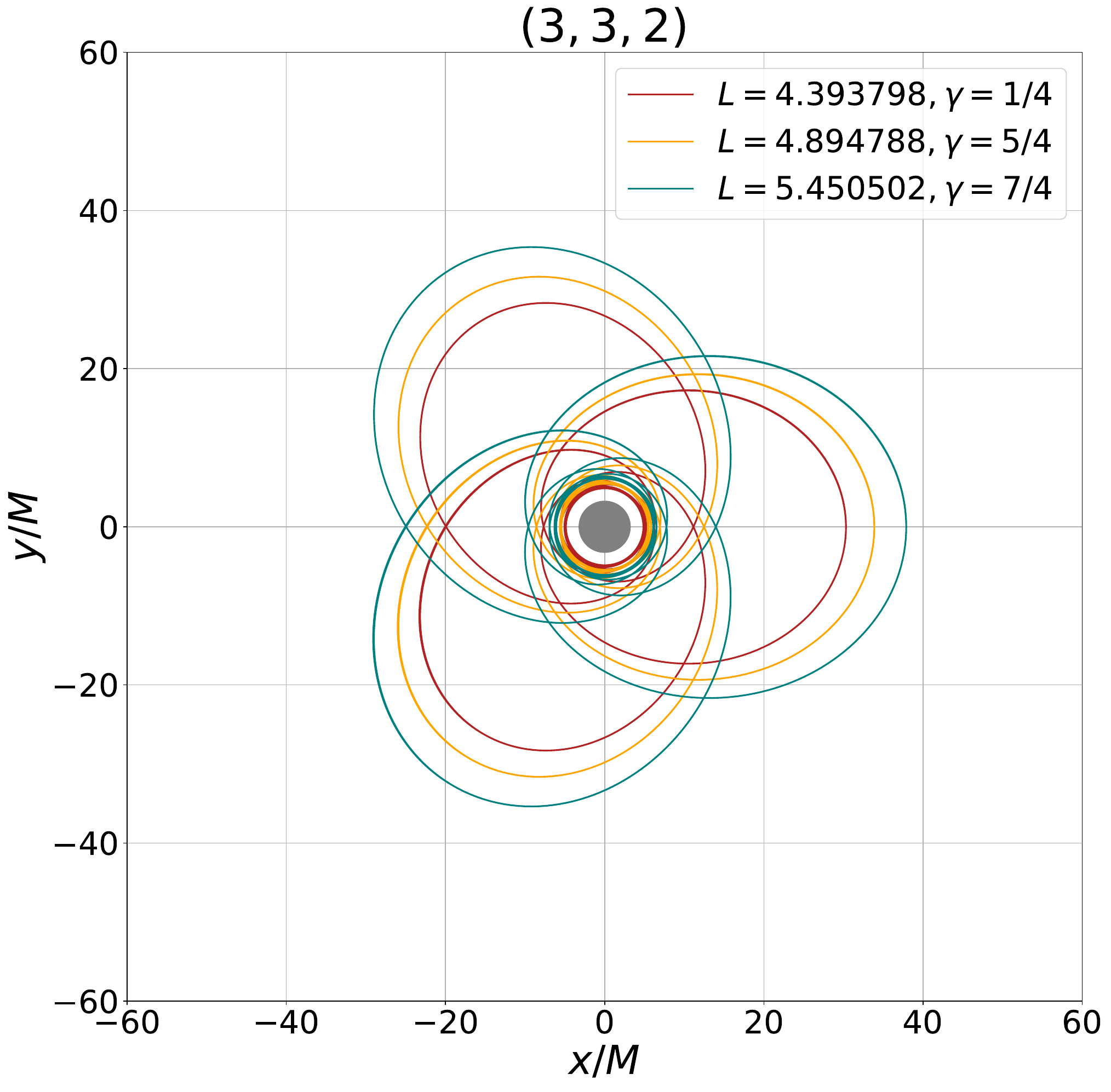} \\
    
    \vspace{0.2cm} 
    \includegraphics[width=0.32\textwidth]{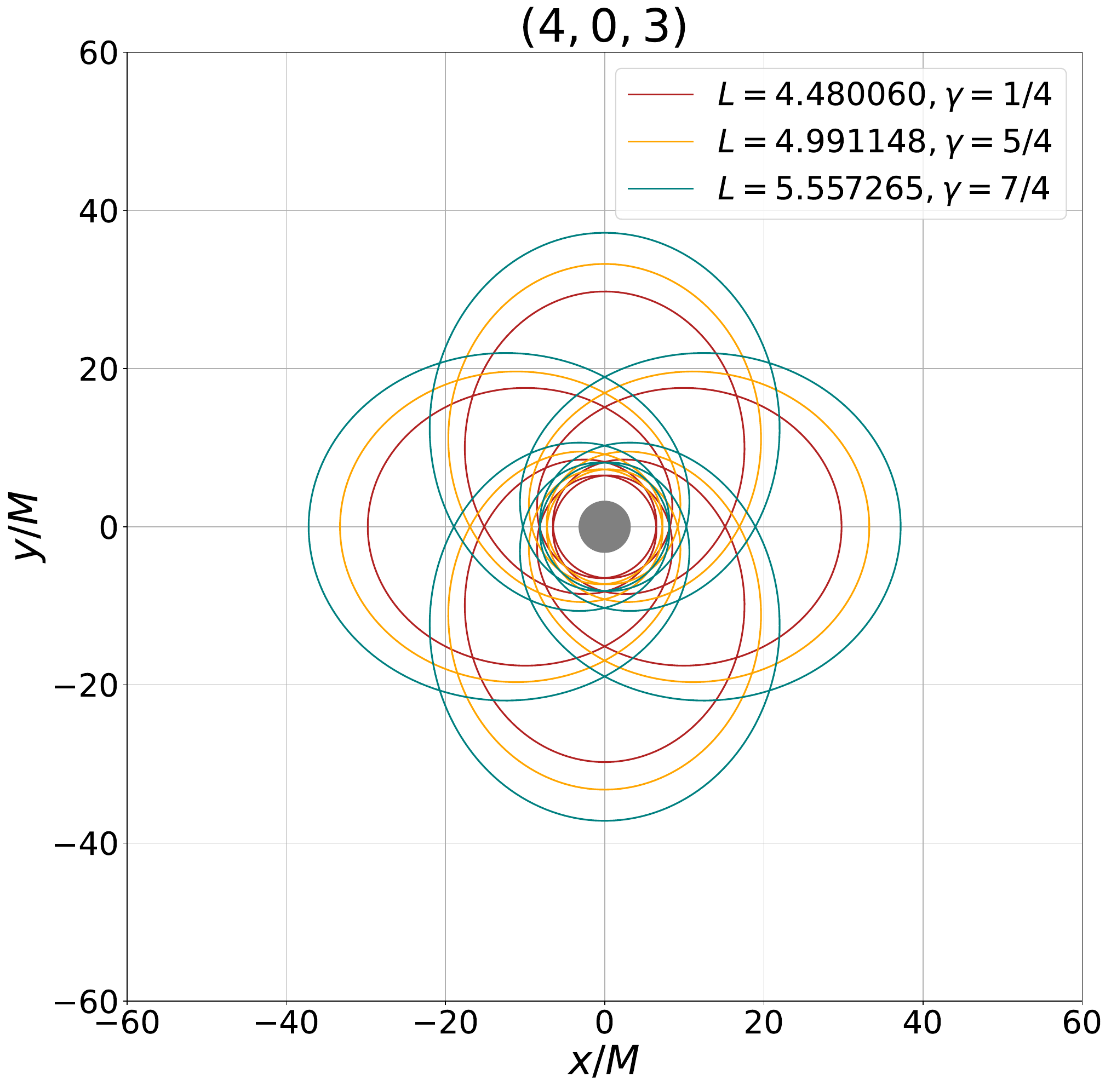} \hfill
    \includegraphics[width=0.32\textwidth]{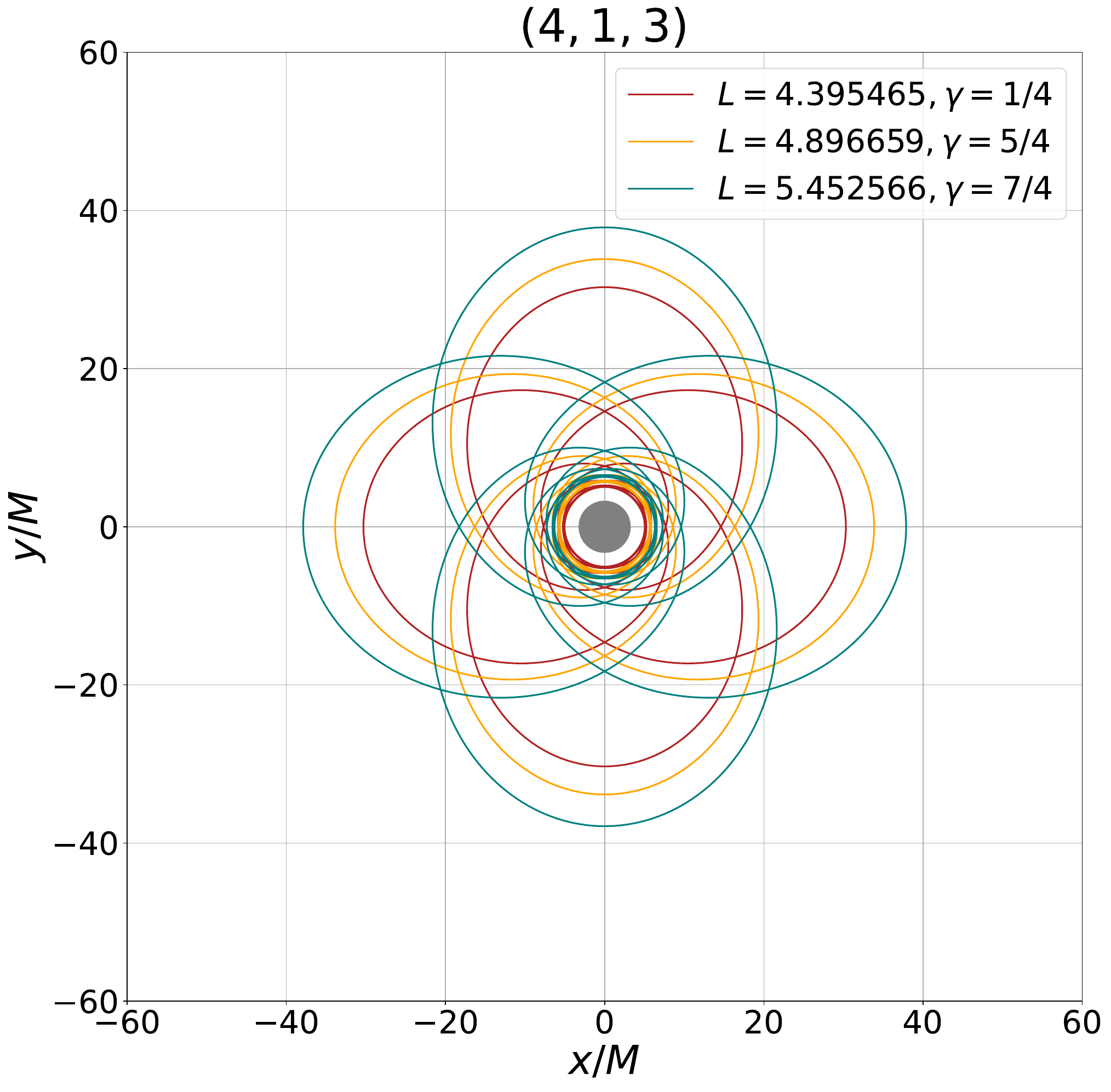} \hfill
    \includegraphics[width=0.32\textwidth]{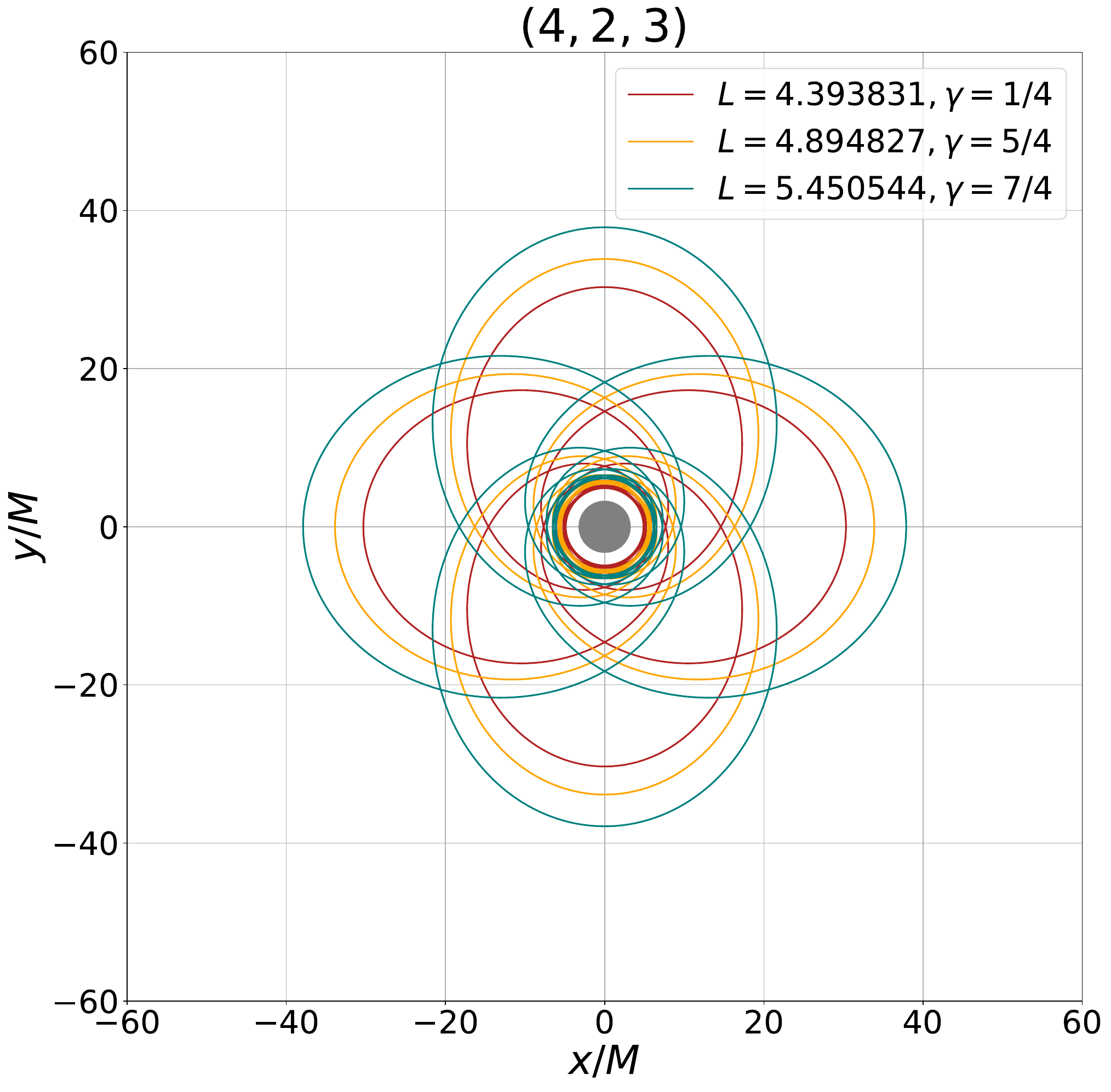}
    
    \caption{Periodic orbits around a generalized Schwarzschild-like BH surrounded by a Dehnen-type DM halo are shown for various combinations of $(z,w,v)$ and $\gamma$. In all cases, we set $\rho_s=0.4$, $r_s=0.5$, and $E=\frac{1}{2}(E_{MBO}+E_{ISCO})$.}
    \label{fig:periodicL}
\end{figure*}
\begin{figure*}[htbp]
\includegraphics[width=1\textwidth]{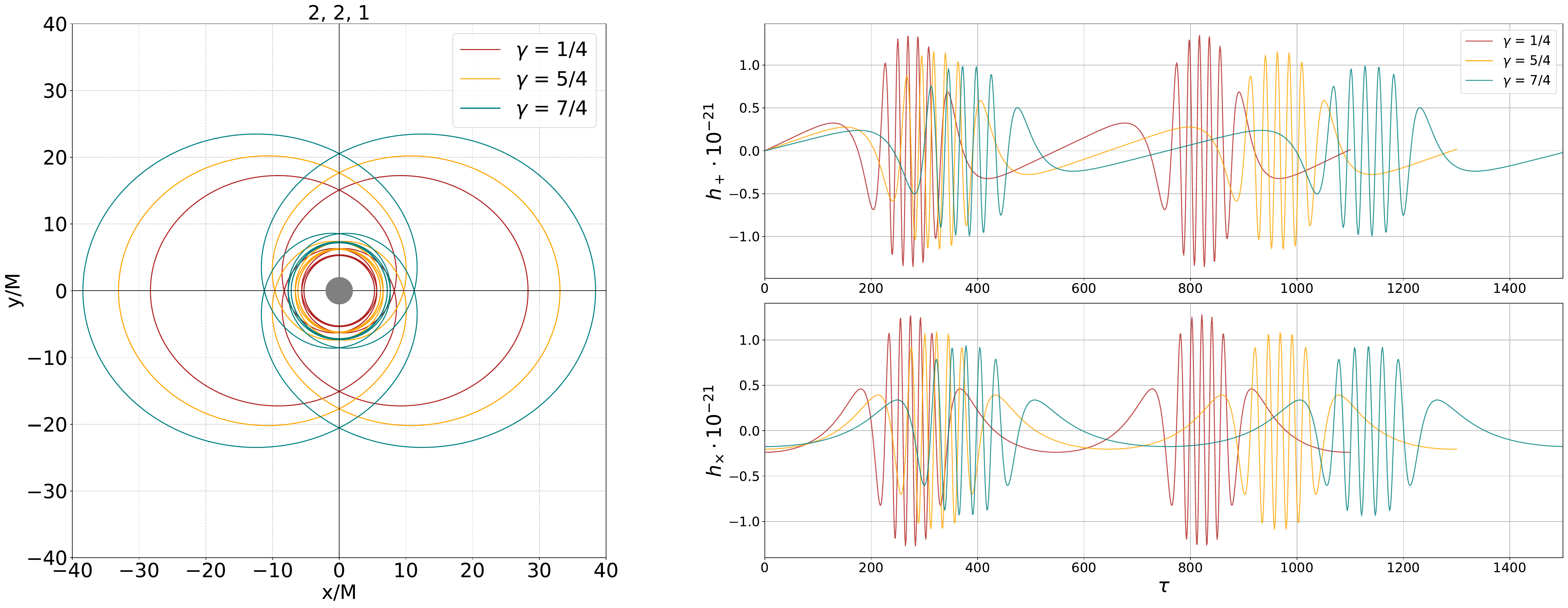}
\includegraphics[width=1\textwidth]{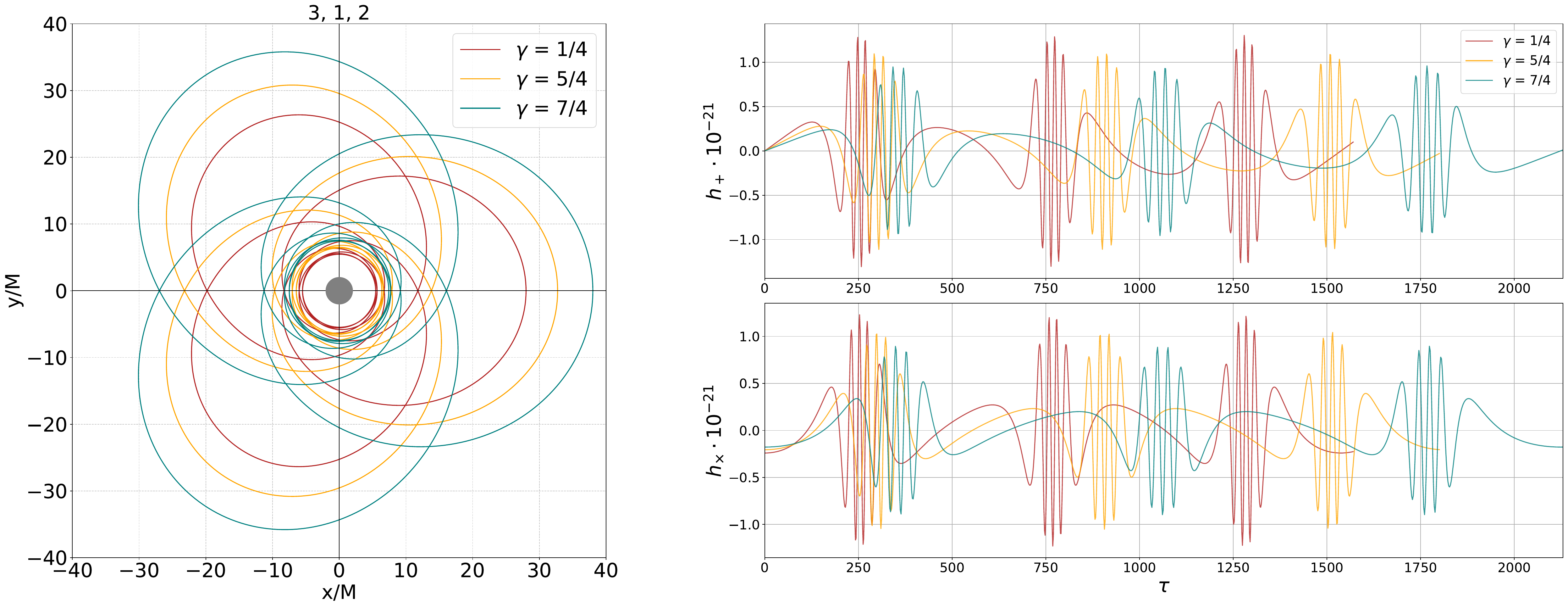}
\includegraphics[width=1\textwidth]{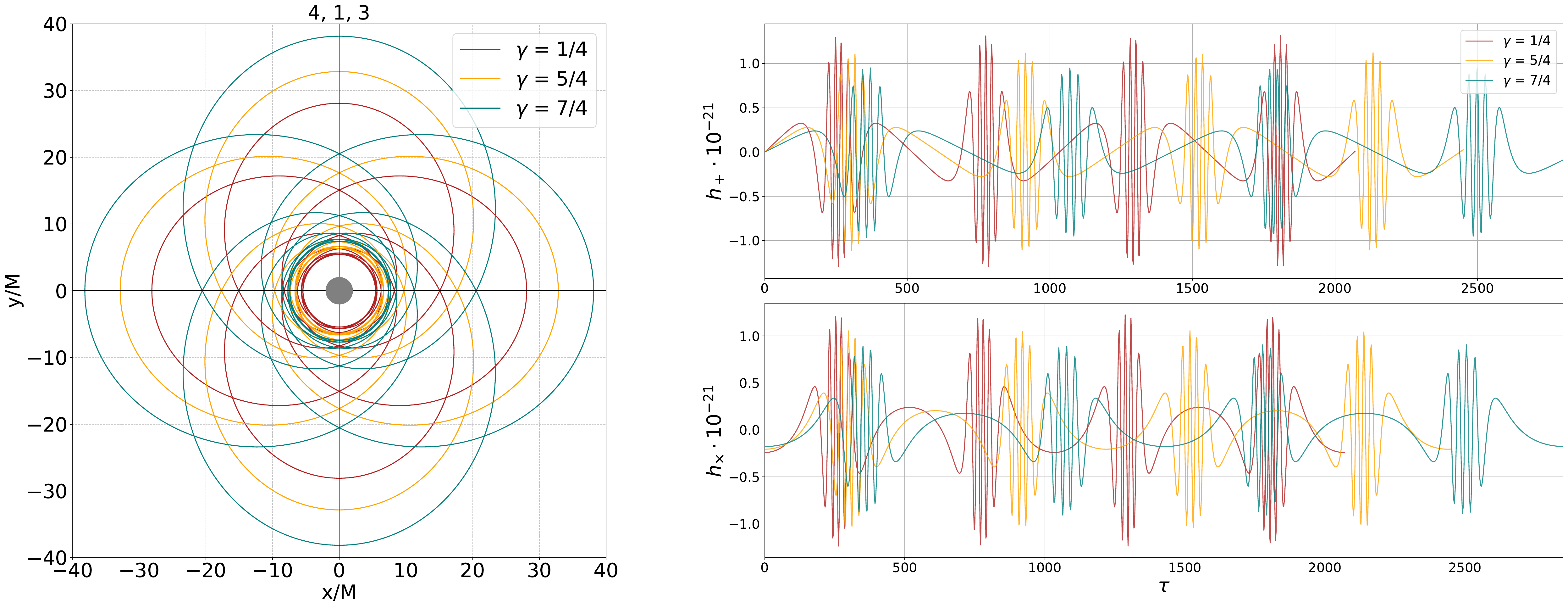}
\caption{Periodic orbits and the corresponding gravitational-wave polarizations for an EMRI system consisting of a stellar-mass compact object orbiting a supermassive generalized Schwarzschild-like BH embedded in a Dehnen-type DM halo. From top to bottom, the rows correspond to the $(2,2,1)$, $(3,1,2)$, and $(4,1,3)$ periodic orbits, respectively. The left panels show the orbital trajectories for different values of the halo parameter $\gamma$, while the right panels display the corresponding $h_{+}$ and $h_{\times}$ waveforms. We set $r_s=0.4$, $\rho_s=0.4$, $M\sim10^{6}M_{\odot}$, and $m\sim10M_{\odot}$.
}
\label{fig:GW}
\end{figure*}
\begin{figure*}[htbp]
\includegraphics[width=1\textwidth]{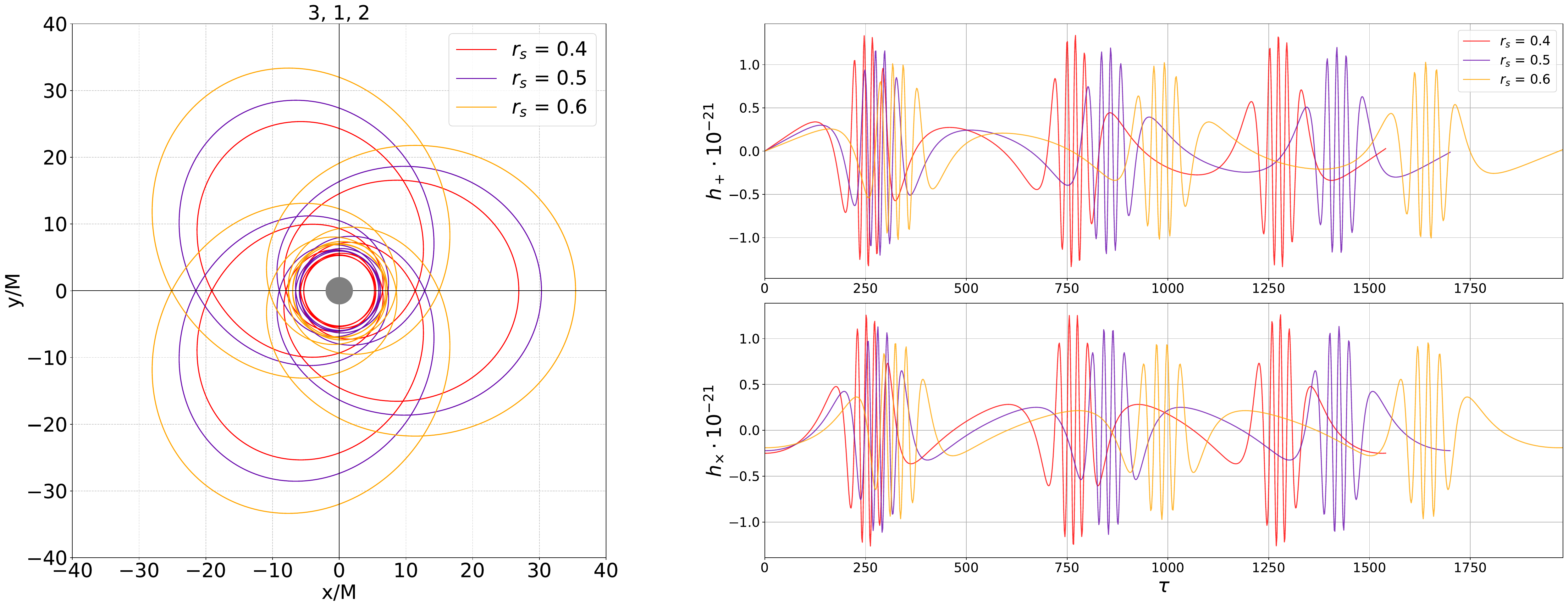}
\includegraphics[width=1\textwidth]{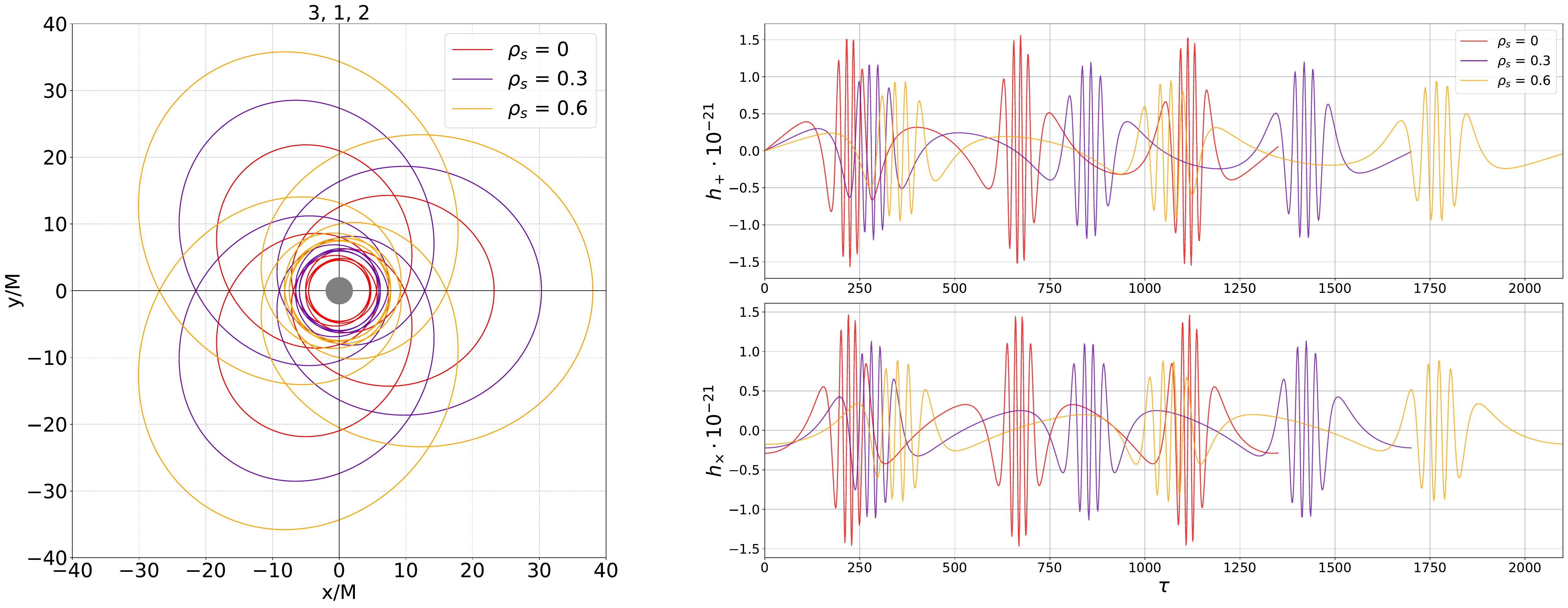}
\caption{The $(3,1,2)$ periodic orbit and the corresponding gravitational-wave polarizations for an EMRI system consisting of a stellar-mass compact object ($m\sim10M_{\odot}$) orbiting a supermassive generalized Schwarzschild-like BH ($M\sim10^{6}M_{\odot}$) embedded in a Dehnen-type DM halo. The upper row shows the effect of varying the scale radius $r_s$ at fixed $\rho_s=0.4$, while the lower row illustrates the effect of varying the halo density $\rho_s$ at fixed $r_s=0.4$, with $\gamma=7/4$ in both cases. In each row, the left panel displays the orbital trajectories, while the two right panels show the corresponding $h_{+}$ and $h_{\times}$ waveforms.}
\label{fig:GW2}
\end{figure*}

\section{Periodic orbits around the Schwarzschild BH surrounded by DM halo}\label{sec3}

In the previous section, we considered the impact of the DM halo parameters on timelike geodesics around the generalized Schwarzschild-like BH in a Dehnen-type DM halo. In this section, we explore the periodic orbits around the above mentioned BH spacetime. Following Levin and Perez-Giz~\cite{Levin_2008}, we introduce the periodic orbits through a triplet of integers ($z,w,v$) that correspond to the zoom, whirl, and vertex behavior of the trajectory, respectively. In this scenario, periodic orbits can be modeled as follows: a massive particle returns to its initial position after a finite time and a finite number of radial ${\omega_{r}}$ and angular ${\omega_{\varphi}}$ oscillations, requiring that the ratio of these orbital frequencies is a rational number. Here, the rational number is given by
\begin{equation} \label{eq:rational_q}
    q=\frac{\omega_{\phi}}{\omega_r}-1=w+\frac{v}{z}\, .
\end{equation}
The equations of motion allows one to write the rational number as follows \cite{2025JCAP...01..091Y,SHABBIR2025101816,JIANG2024101627}:
\begin{widetext}
\begin{equation} \label{eq:periodic_q}
    q=\frac{1}{\pi}\int^{r_2}_{r_1} \frac{\dot{\phi}}{\dot{r}}-1=\frac{1}{\pi} \int^{r_2}_{r_1}\frac{Ldr}{r^2\sqrt{E^2-\Big[1-\frac{2M}{r}-\frac{8\pi\rho_sr_s^3}{(3-\gamma)r}\Big(\frac{r}{r+r_s}\Big)^{3-\gamma}\Big]\Big(1+\frac{L^2}{r^2}\Big)}} - 1\, ,
\end{equation}
\end{widetext}
with $r_1$ and $r_2$ being the periapsis and apoapsis radii of the periodic orbits.

We now investigate the dependence of the rational number $q$ on the particle energy $E$ and orbital angular momentum $L$ for different values of the Dehnen-type DM halo parameter $\gamma$, as shown in Fig.~\ref{fig:q}. The left panel shows the variation of $q$ with respect to $E$ for fixed angular momentum $L=(L_{\rm MBO}+L_{\rm ISCO})/2$. It is observed that $q$ increases with the particle energy and rises sharply as $E$ approaches the upper boundary of the allowed energy interval.
 In addition, increasing $\gamma$ shifts the curves toward larger values of $E$, indicating that a higher particle energy is required to obtain the same rational number for larger values of $\gamma$.
The right panel shows the behavior of $q$ as a function of the orbital angular momentum $L$ for fixed energy $E=0.96$. In this case, $q$ increases as $L$ approaches its lower limiting value and then decreases gradually as $L$ increases.
 Moreover, larger values of $\gamma$ shift the curves toward higher angular momenta. This demonstrates that the halo profile parameter $\gamma$ modifies the allowed ranges of energy and angular momentum associated with periodic orbits.

To construct the periodic trajectories, we numerically solve Eqs.~\eqref{eq:rational_q} and \eqref{eq:periodic_q} for different choices of the triplet $(z,w,v)$.  First, we fix the angular momentum as
$L=(L_{\rm MBO}+L_{\rm ISCO})/2$ and compute the corresponding energy required for each periodic-orbit configuration. The results are summarized in Table~\ref{table1}. Next, by fixing the particle energy at $E=(E_{\rm MBO}+E_{\rm ISCO})/2$, we determine the angular momentum associated with the same set of periodic-orbit configurations. The corresponding results are presented in Table~\ref{table2}. Based on the numerical values listed in Tables~\ref{table1} and~\ref{table2}, we construct the corresponding periodic trajectories around the generalized Schwarzschild-like BH embedded in a Dehnen-type DM halo. The resulting trajectories are shown for different orbital configurations $(z,w,v)$ and different values of the halo parameter $\gamma$ in Figs.~\ref{fig:periodic} and~\ref{fig:periodicL}.

As can be seen from Figs.~\ref{fig:periodic} and~\ref{fig:periodicL}, the structure of the periodic orbits is governed by the integers $(z,w,v)$. Increasing the zoom number $z$ produces more complex orbital patterns with a larger number of leaves, whereas increasing the whirl number $w$ leads to more revolutions around the central BH before the particle reaches the next apoapsis. Moreover, for a fixed orbital configuration $(z,w,v)$, larger values of the Dehnen-type halo parameter $\gamma$ enlarge the periodic trajectories and shift them farther from the BH. This shows that the halo profile parameter $\gamma$ plays an important role in modifying the structure of periodic orbits.
\begin{figure*}[htbp]
\centering
 \includegraphics[scale=0.3]{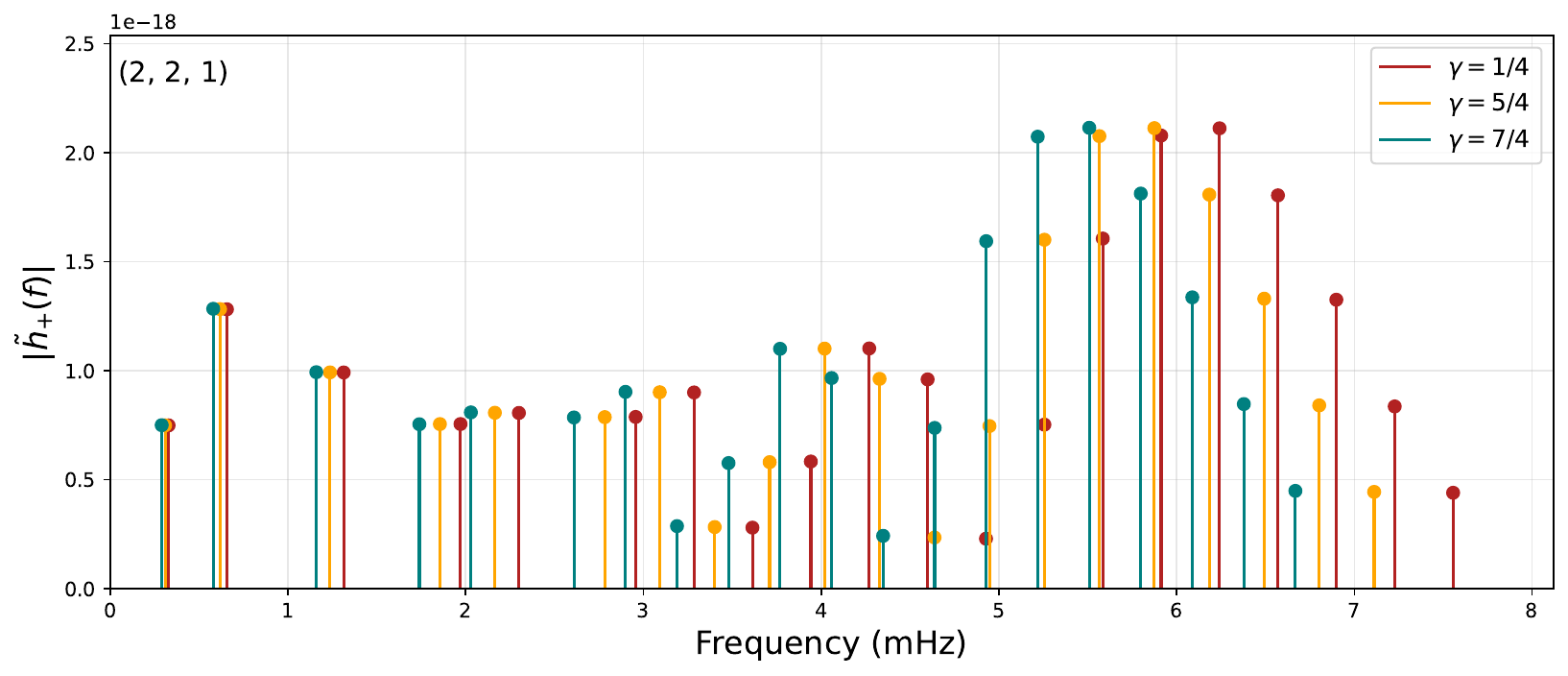}
 \includegraphics[scale=0.3]{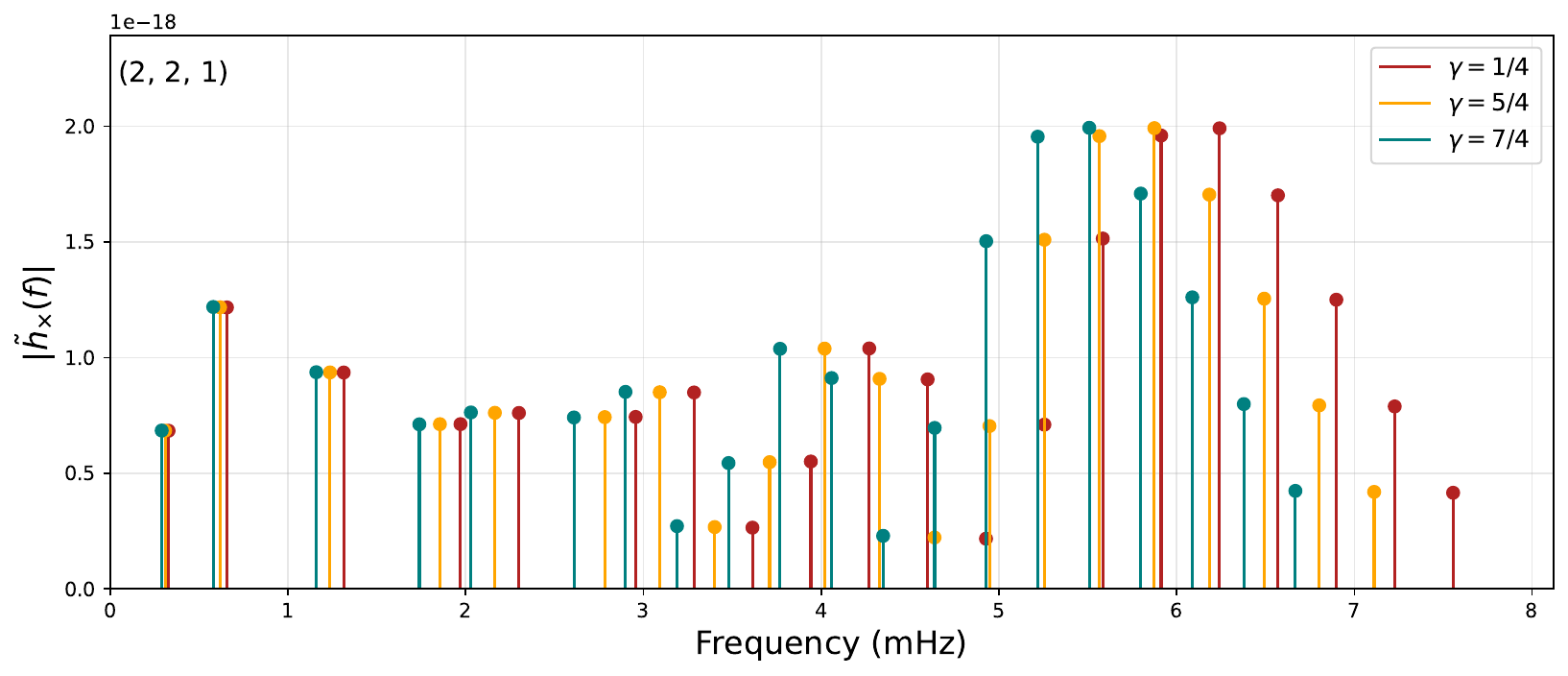}
 \includegraphics[scale=0.3]{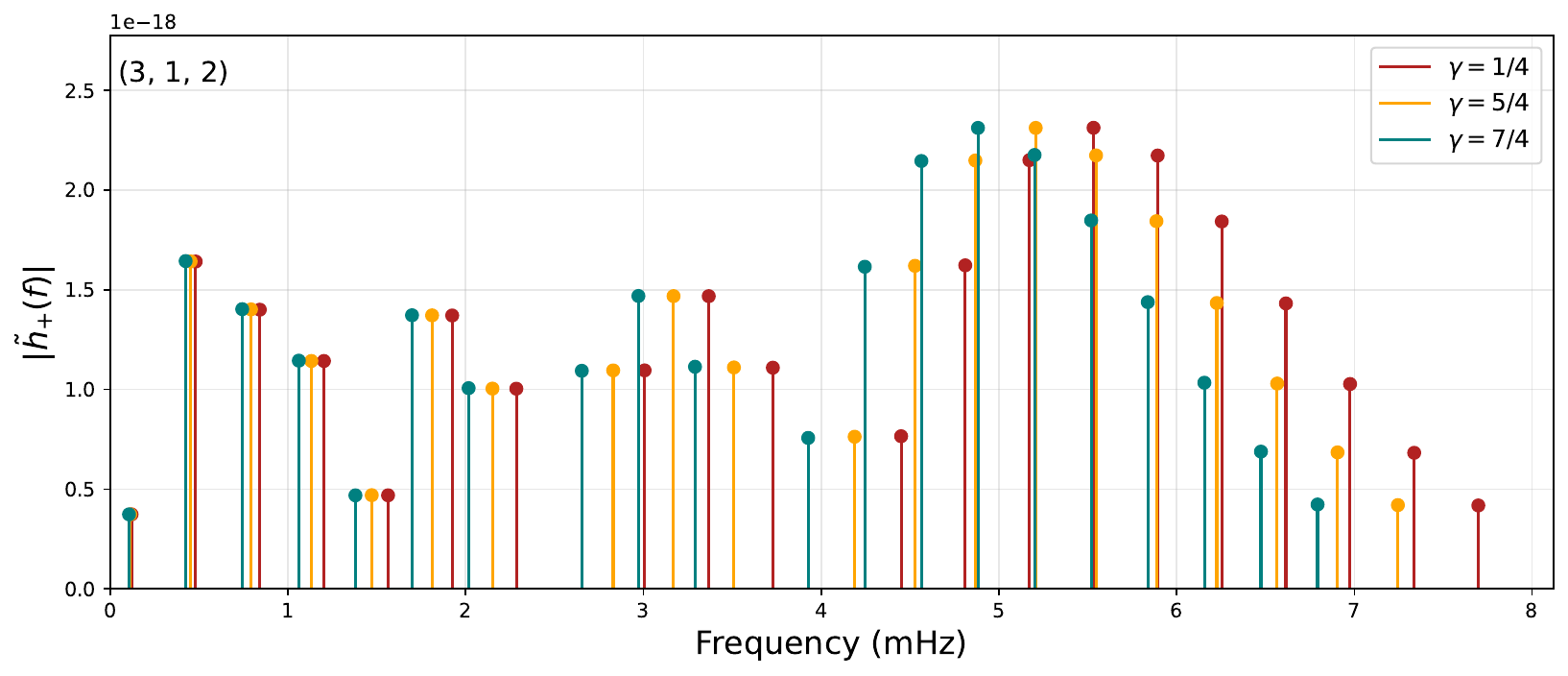}
 \includegraphics[scale=0.3]{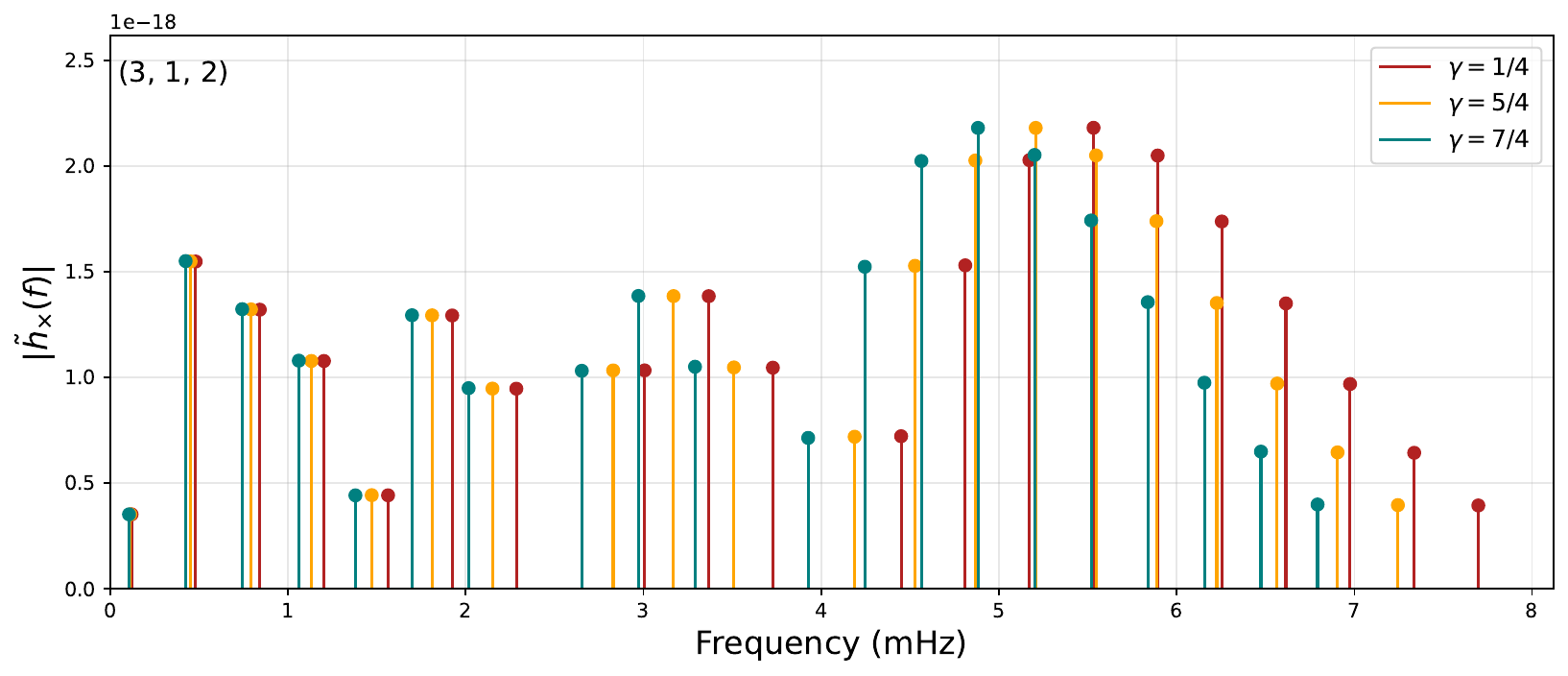}
  \includegraphics[scale=0.3]{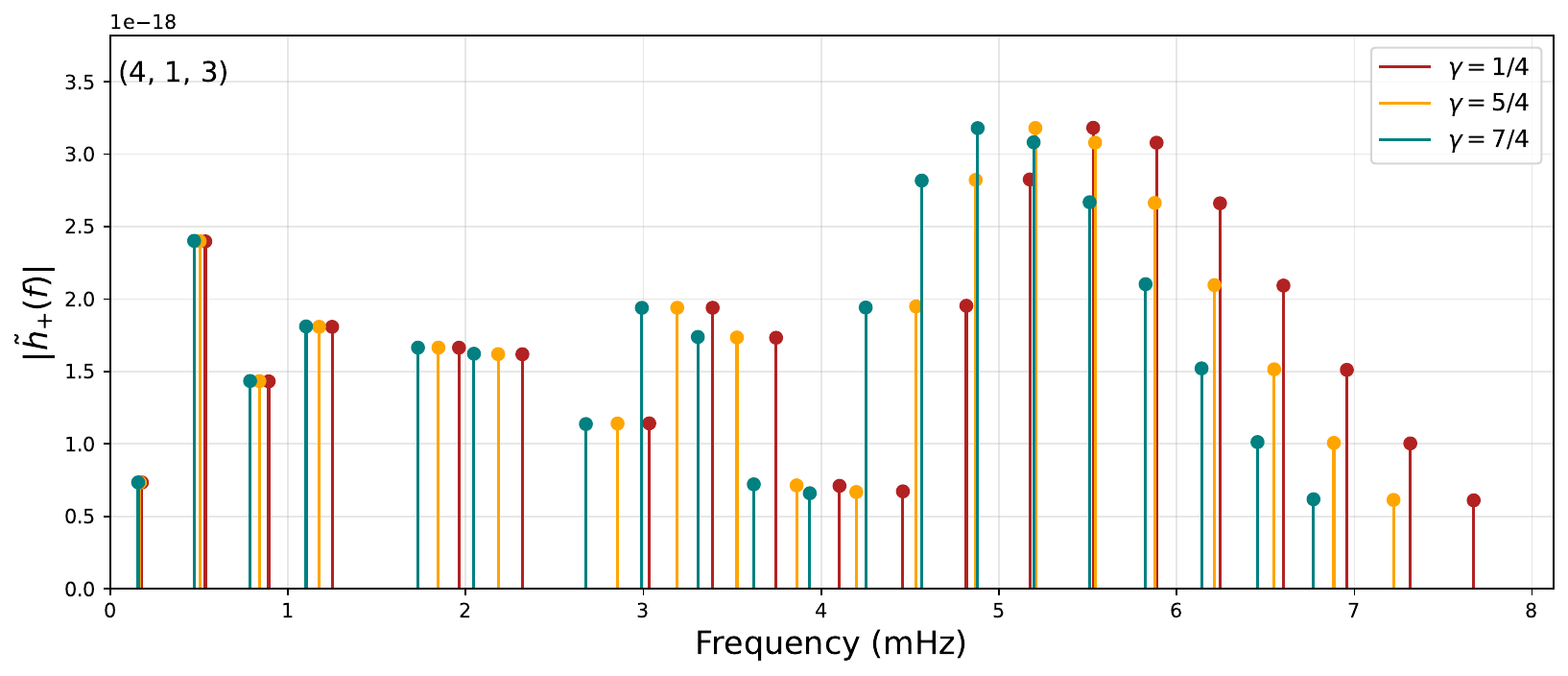}
 \includegraphics[scale=0.3]{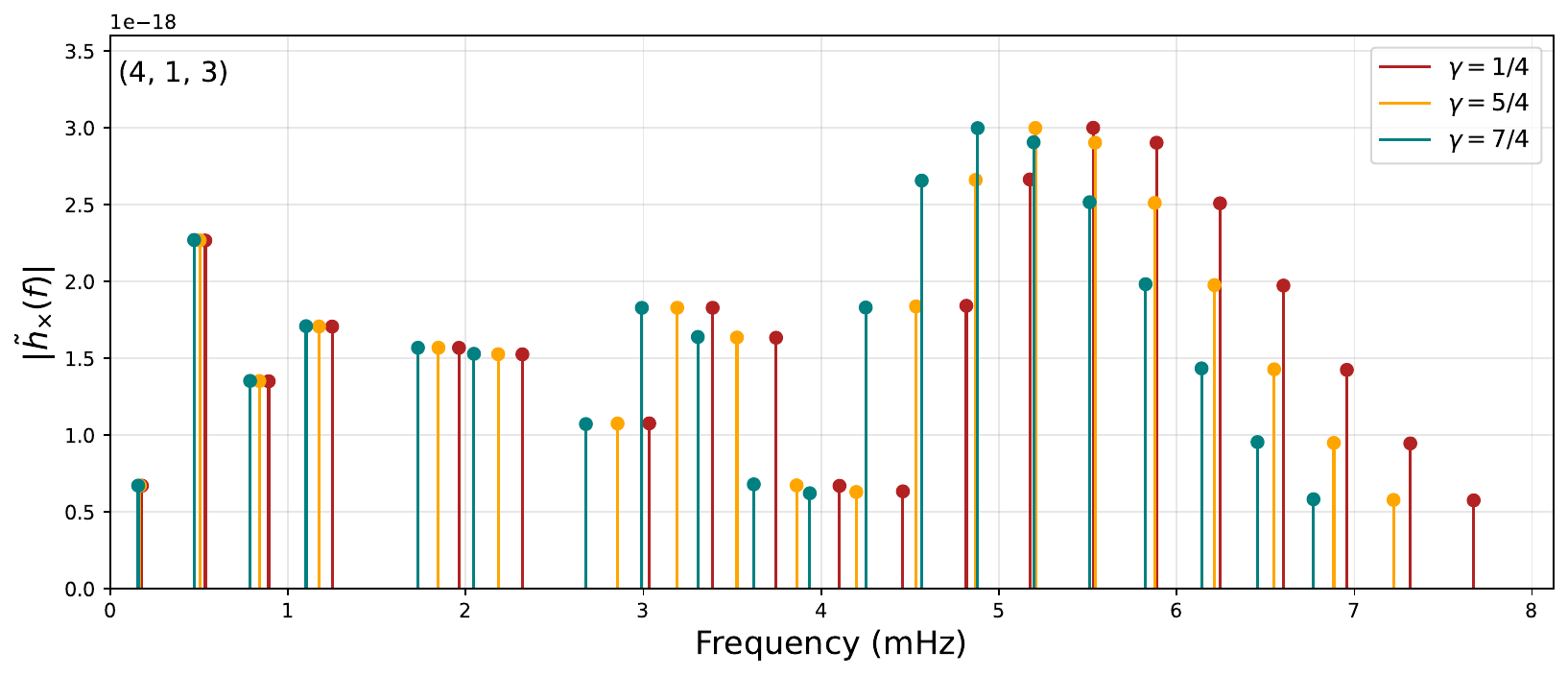}
 \caption{Frequency spectra of the $h_{+}$ and $h_{\times}$ gravitational-wave polarizations for the $(2,2,1)$, $(3,1,2)$, and $(4,1,3)$ periodic orbits at different values of the halo parameter $\gamma$.}
 \label{fig:spectrum}
\end{figure*}
\begin{figure*}[htbp]
\centering
 \includegraphics[scale=0.27]{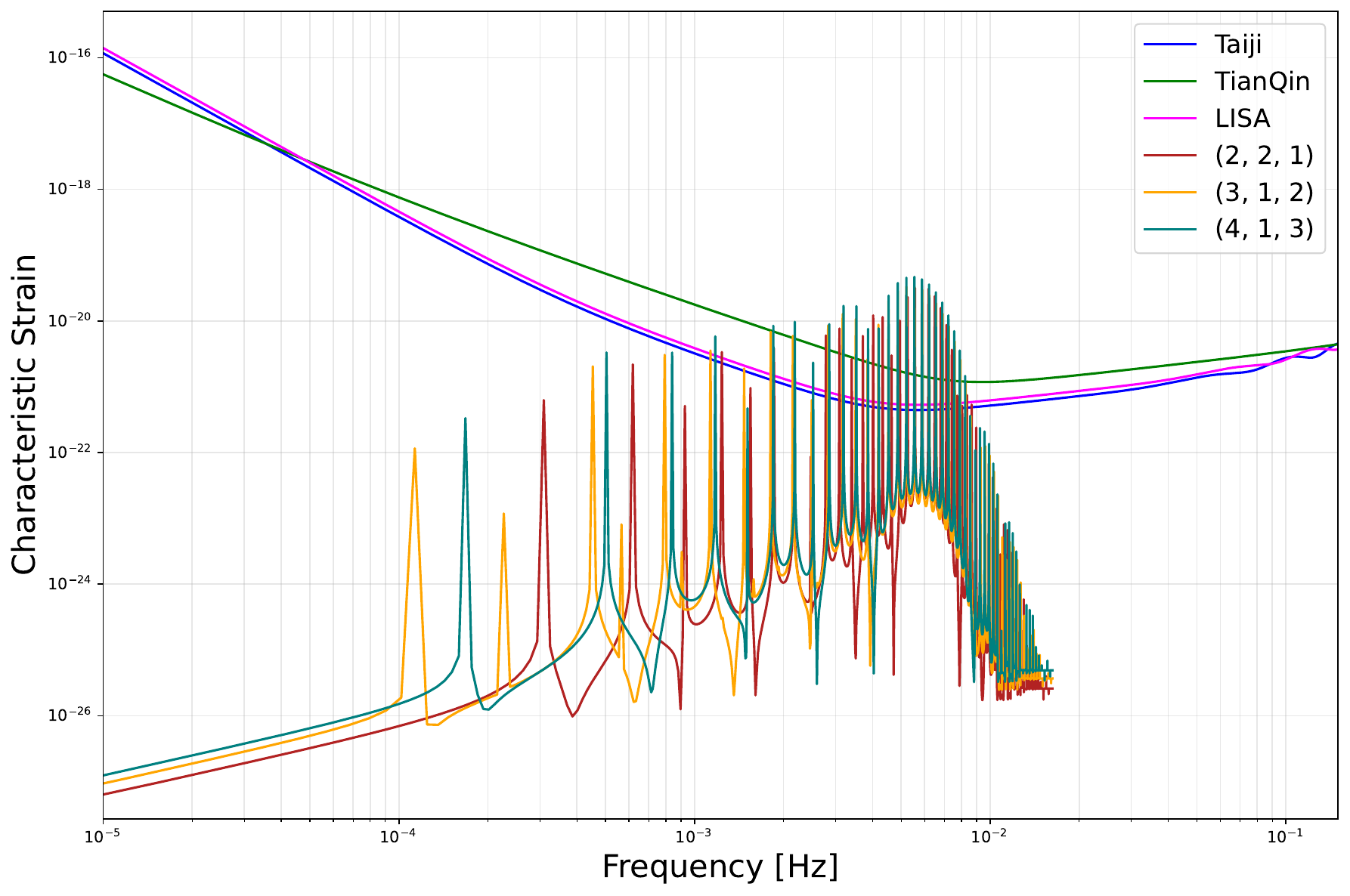}
 \includegraphics[scale=0.27]{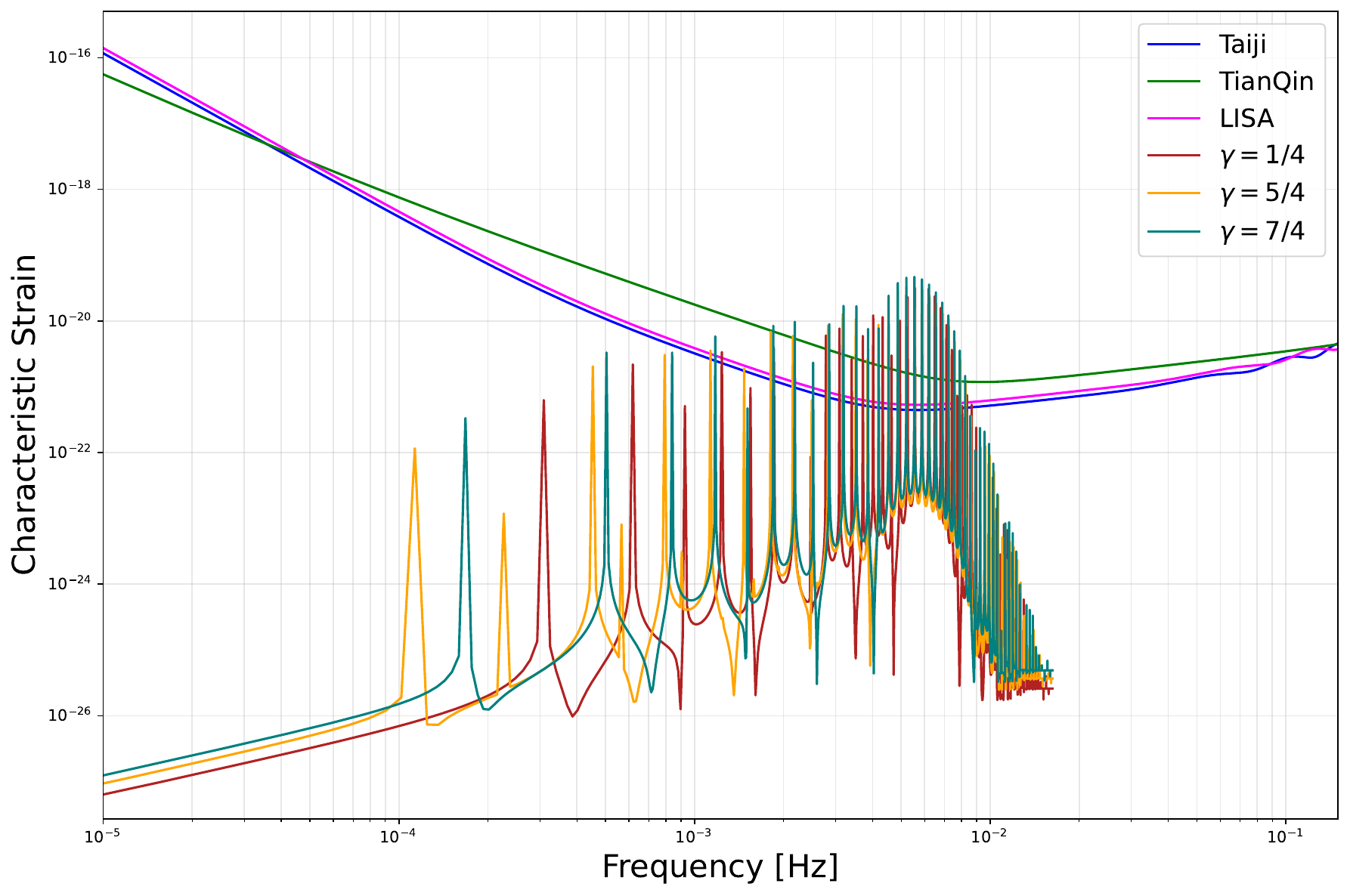}
 \caption{Characteristic strains of the gravitational-wave signals compared with the sensitivity curves of LISA, Taiji, and TianQin. The left panel corresponds to different periodic orbits at fixed $\gamma=5/4$, whereas the right panel shows the $(3,1,2)$ orbit for different values of $\gamma$.
}
 \label{fig:strain}
\end{figure*}
 
\section{NUMERICAL KLUDGE GRAVITATIONAL WAVEFORMS FROM PERIODIC ORBITS}\label{sec4}

Next-generation space-based gravitational-wave observatories, including LISA, Taiji, and TianQin, are expected to open new possibilities for detecting and measuring gravitational waves with high precision. Among the most important target sources for these detectors are mergers of supermassive BHs in galactic centers, compact stellar binaries, and extreme mass-ratio inspirals (EMRIs), in which a stellar-mass compact object orbits a massive BH. The gravitational waves emitted from such an EMRI system can encode information about the periodic orbital dynamics and the spacetime geometry of the central supermassive BH. Motivated by this, we consider an EMRI system in which a stellar-mass compact object orbits a supermassive BH described by the generalized Schwarzschild-like spacetime embedded in a Dehnen-type DM halo.

To model the gravitational waveforms, we adopt the adiabatic approximation method. In this framework, the inspiral timescale is assumed to be much longer than the orbital period; hence, the energy and angular momentum of the smaller object vary slowly and can be treated as approximately conserved over many orbital cycles. Consequently, the motion of the smaller object can be described as a sequence of geodesics in the background spacetime.

We use the numerical kludge waveform model to compute the gravitational waveforms emitted by a compact object moving along periodic orbits around the central BH. Within this approach, the trajectory $Z^{i}(t)$ of the smaller compact object is first obtained by numerically solving the geodesic equations of motion, Eqs.~\eqref{eq:azimthal} and~\eqref{eq:radialeq}, in the generalized Schwarzschild-like BH spacetime embedded in a Dehnen-type DM halo. Once the orbital trajectory $Z^{i}(t)$ has been determined, it is substituted into the quadrupole radiation formula to construct the corresponding gravitational waveform. This method provides a computationally efficient approximation to EMRI signals and allows us to investigate their sensitivity to the orbital configuration and the geometry of the background spacetime. In the quadrupole formalism, the gravitational-wave metric perturbation $h_{ij}$ generated by a symmetric trace-free (STF) mass quadrupole moment $I_{ij}$ is given by \cite{2025JCAP...01..091Y}
\begin{equation}
h_{ij}=\frac{2G}{c^4D_L},\ddot{I}_{ij},
\end{equation}
where $D_L$ denotes the luminosity distance to the source. The STF mass quadrupole moment associated with a point-like compact object of mass $m$ takes the form
\begin{equation}
I^{ij}
=
\left[
\int d^{3}x \, x^{i} x^{j} \, T^{tt}(t,x^{i})
\right]^{\mathrm{STF}}\, ,
\end{equation}
where $T^{tt}$ is the $tt$ component of the stress-energy tensor for a point-like compact object following the trajectory $Z^{i}(t)$,
\begin{equation}
T^{tt}(t,x^{i}) = m \, \delta^{3}\!\left(x^{i} - Z^{i}(t)\right)\, .
\end{equation}
Since the geodesic equations and the corresponding particle trajectories are expressed in Boyer–Lindquist coordinates, the quadrupole moment of the EMRI system ($M \gg m$) takes the  ~\cite{2025EPJC...85...36Z}
\begin{equation}\label{eq:perturbation}
h_{ij}=\frac{4\eta M}{D_{L}}\left(v_{i}v_{j}-\frac{M}{r}n_{i}n_{j}\right)\, .
\end{equation}
where $M$ and $m$ denote the masses of the central BH and the point-like compact object, respectively, while $\eta =Mm/(M+m)^2$ is the symmetric mass ratio of the EMRI system. Furthermore, $r$ denotes the instantaneous radial distance between the smaller object and the central BH, determined from the numerical solution of the geodesic equations of motion. The quantities $v_i$ and $n_i$ are the $i$th components of the smaller object's orbital velocity and of the radial unit vector pointing from the BH toward the smaller object, respectively.

To express the gravitational-wave signal in a form suitable for analysis, we introduce a Cartesian frame $(x,y,z)$ centered on the BH. The transformation from Boyer--Lindquist coordinates is given by
\begin{equation} 
x = r \sin\theta \cos\phi, \quad
y = r \sin\theta \sin\phi, \quad
z = r \cos\theta\,  .
\end{equation}
We now introduce a detector-adapted coordinate system $(X,Y,Z)$ that shares the same origin as the Cartesian frame $(x,y,z)$ ~\cite{Poisson_Will_2014,2025JCAP...01..091Y,2025EPJC...85...36Z}. The detector-frame basis vectors can be expressed in terms of the basis vectors of the original Cartesian frame as 
\begin{equation}\label{xyz}
\begin{array}{l}
e_{X}=[\cos\zeta,-\sin\zeta,0]\, ,\\
e_{Y}=[\cos\iota\sin\zeta,\cos\iota\cos\zeta,-\sin\iota]\, ,\\
e_{Z}=[\sin\iota\sin\zeta,\sin\iota\cos\zeta,\cos\iota]\, .
\end{array}
\end{equation}
where $\iota$ denotes the orbital inclination relative to the detector $X–Y$ plane, while $\zeta$ specifies the longitude of the pericenter in the orbital plane. Using the above relations, the plus and cross polarizations of the gravitational wave are obtained as~\cite{2025EPJC...85...36Z,2024arXiv241101858M,2025JCAP...01..091Y}:
\begin{equation} \label{eq:hplus}
 h_{+}=\frac{1}{2}(e^i_X e^j_X-e^i_Ye^j_Y)h_{ij},
\end{equation}
\begin{equation} \label{eq:hcross}
     h_{\times}=\frac{1}{2}(e^i_Xe^j_Y+e^i_Ye^j_X)h_{ij}
\end{equation}
Combining Eqs.~\eqref{eq:perturbation}, \eqref{xyz}, \eqref{eq:hplus}, and~\eqref{eq:hcross}, we obtain the following explicit expressions for the gravitational-wave polarizations $h_{+}$ and $h_{\times}$ ~\cite{2025EPJC...85...36Z,2024arXiv241101858M}:
\begin{equation} \label{eq:h_plus}
    h_{+}=-\frac{2\mu M^{2}}{D_{L}r}\left(1+\cos^{2}\iota\right)\cos\left(2\phi+2\zeta\right)\, ,
\end{equation}
\begin{equation}\label{eq:h_cross}
    h_{\times}=-\frac{4\mu M^{2}}{D_{L}r}\cos\iota\sin\left(2\phi+2\zeta\right)\, ,
\end{equation}
where the coordinates $(r,\phi)$ determined through numerical integration of the geodesic equations in the Boyer--Lindquist coordinates.

To obtain the gravitational-wave signals from the periodic trajectories, we model the source as an EMRI in which a $10M_{\odot}$ compact object orbits a $10^{6}M_{\odot}$ supermassive BH. The orbital orientation is fixed by setting both $\zeta$ and $\iota$ to $\pi/4$, while the source is assumed to lie at a luminosity distance of $D_L=200\,\mathrm{Mpc}$.

Fig.~\ref{fig:GW} shows the gravitational-wave polarizations $h_{+}$ and $h_{\times}$ emitted over one complete orbital period by the $(2,2,1)$, $(3,1,2)$, and $(4,1,3)$ periodic orbits for different values of the Dehnen-type DM halo parameter $\gamma$. The waveforms clearly exhibit distinct zoom and whirl phases. The smooth, slowly varying portions of the signal correspond to the zoom phase, during which the compact object moves through the outer region of the orbit. In contrast, the rapidly oscillating segments are radiated during the whirl phase, when the object approaches the central BH and follows a nearly circular trajectory around it. From the trajectories shown in the left panels, it can be seen that increasing $\gamma$ enlarges the radial extent of the periodic orbits. As a result, the orbital period increases, the amplitudes of the polarization components decrease.
Moreover, periodic orbits with a larger zoom number $z$ exhibit longer periods and a greater number of recurring waveform patterns. 

Fig.~\ref{fig:GW2} shows how the halo scale radius $r_s$ and density $\rho_s$ affect the $(3,1,2)$ periodic orbit and its gravitational-wave signal. Increasing either $r_s$ or $\rho_s$ produces an effect similar to that of increasing $\gamma$: the orbit expands, the orbital period becomes longer, the polarization amplitudes decrease, and the high-frequency waveform segments become more clearly separated.

To examine the spectral content of the emitted gravitational radiation, we transform the time-domain waveforms in Fig.~\ref{fig:GW} into the frequency domain using a discrete Fourier transform. The corresponding spectra, $\tilde{h}_{+}(f)$ and $\tilde{h}_{\times}(f)$, are displayed in Fig.~\ref{fig:spectrum}. The dominant spectral features of all three periodic-orbit configurations occur in the millihertz band, which falls within the observational range of future space-based gravitational-wave detectors.
The figure further shows that increasing the Dehnen-type halo parameter $\gamma$ shifts the spectral lines toward lower frequencies. This shift is more noticeable in the high-frequency part of the spectrum, while the low-frequency components show smaller shifts. Such behavior is consistent with the increase in the orbital period caused by larger values of $\gamma$.

The characteristic gravitational-wave strain is defined as
\begin{equation}\label{eq}
h_c(f)=2f\left(
\left|\widetilde{h}_{+}(f)\right|^2+
\left|\widetilde{h}_{\times}(f)\right|^2
\right)^{1/2}.
\end{equation}
The characteristic strains obtained from Eq.~\eqref{eq} are shown in Fig.~\ref{fig:strain}, together with the sensitivity curves of LISA, Taiji, and TianQin. The left panel compares different periodic-orbit configurations, while the right panel illustrates the influence of the halo parameter $\gamma$. A sequence of spectral peaks appears in the millihertz range and reflects the effects of the orbital dynamics and the surrounding DM distribution. Several of these peaks lie above the detector sensitivity curves, suggesting that the DM halo may leave observable imprints on EMRI gravitational-wave signals.

\section{Summary}\label{summary}

In this study, we examined how the DM halo influences the geodesic motion of test particles around a Schwarzschild-like BH embedded in a generalized Dehnen-type DM halo. The effective-potential analysis showed that increasing the density profile parameter $\gamma$ shifts the unstable circular orbit inward and the stable circular orbit outward, indicating that this halo parameter affects the orbital configurations of test particles. We also found that larger values of $\rho_s$ and $\gamma$ increase the radii and angular momenta of the MBO and ISCO. By contrast, the ISCO energy decreases as $r_s$ and $\rho_s$ increase. These results demonstrate that the DM halo modifies both the orbital structure, the angular momentum and the energy of particles at the ISCO.

We also studied how the density profile parameter $\gamma$ affects the dependence of the rational number $q$ on the energy $E$ and orbital angular momentum $L$. We found that larger values of $\gamma$ shift the $q(E)$ and $q(L)$ curves toward higher values of energy and angular momentum, respectively. These findings indicate that $\gamma$ modifies the ranges of energy and angular momentum associated with periodic orbits. We further calculated the energy and angular momentum corresponding to various periodic-orbit configurations $(z,w,v)$. The numerical results were tabulated in Tables~\ref{table1} and~\ref{table2} and used to generate the associated periodic trajectories.

We next investigated the gravitational-wave signatures associated with periodic orbits in the EMRI system. Specifically, we considered a stellar-mass compact object with $m\sim10M_{\odot}$ orbiting a supermassive generalized Schwarzschild-like BH with $M\sim10^{6}M_{\odot}$ embedded in a Dehnen-type DM halo. We obtained the gravitational-wave signals corresponding to the $(2,2,1)$, $(3,1,2)$, and $(4,1,3)$ periodic orbits within the numerical kludge framework. The results showed that larger values of $\gamma$, $\rho_s$, and $r_s$ produce larger periodic orbits, longer orbital periods, and lower amplitudes for both gravitational-wave polarizations. To further examine the influence of the DM halo on these signals, we analyzed their frequency-domain spectra. The dominant spectral components were found to be in the millihertz range. We also found that, as $\gamma$ increases, the corresponding spectral peaks tend to shift toward lower frequencies. This shift is greater in the high-frequency region and smaller at lower frequencies.

Finally, we analyzed the detectability of the gravitational-wave signals from periodic EMRI orbits. We calculated the characteristic strain for different periodic-orbit configurations $(z,w,v)$ and values of the DM halo parameter $\gamma$. The analysis of the characteristic strain showed that both the orbital configuration and the surrounding DM halo noticeably affect the gravitational-wave signal. Comparison with the sensitivity curves of the future space-based detectors LISA, Taiji, and TianQin revealed that several spectral peaks lie above the detector sensitivity levels. This suggests that the DM halo may leave observable imprints on EMRI gravitational-wave signals.

\bibliographystyle{apsrev4-1}
\bibliography{ref}

\end{document}